\documentclass[twocolumn]{aastex62}

\newcommand{\dd}{{\rm d}}
\newcommand{\Ld}{\mathcal{L}}
\newcommand{\ssb}{\sigma_{\scriptscriptstyle\!\rm SB}}
\newcommand{\Prob}{\mathbb{P}}
\newcommand{\R}{\mathbb{R}}

\newcommand{\myeqref}[1]{Eq.\,(\ref{#1})}
\newcommand{\figref}[1]{Fig.\,\ref{#1}}
\newcommand{\cnc}{55\,Cnc}
\newcommand{\cnce}{55\,Cnc\,e}
\newcommand{\px}{p_\star}
\newcommand{\TD}{TD}
\newcommand{\Tf}{T_{\rm eff}}
\newcommand{\mr}{m_r}

\newcommand{\Zenv}{$Z_{\rm gas}$\xspace}
\newcommand{\Lenv}{$L_{\rm int}$\xspace}
\newcommand{\rc}{$r_{\rm core}$\xspace}
\newcommand{\rsolid}{$r_{\rm core+mantle}$\xspace}
\newcommand{\menv}{$m_{\rm gas}$\xspace}
\newcommand{\fesi}{{\rm Fe}/{\rm Si}_{\rm bulk}\xspace}
\newcommand{\mgsi}{{\rm Mg}/{\rm Si}_{\rm bulk}\xspace}
\newcommand{\fesistar}{{\rm Fe}/{\rm Si}_{\rm star}\xspace}
\newcommand{\mgsistar}{{\rm Mg}/{\rm Si}_{\rm star}\xspace}
\newcommand{\fesima}{{\rm Fe}/{\rm Si}_{\rm mantle}\xspace}
\newcommand{\mgsima}{{\rm Mg}/{\rm Si}_{\rm mantle}\xspace}

\usepackage{soul}
\usepackage{amsmath}
\usepackage{multirow}
\usepackage{xspace}

\received{}
\revised{}
\accepted{19 April 2018}
\submitjournal{ApJ}

\shorttitle{Mass, radius, and composition of the transiting planet \cnce}
\shortauthors{Crida et al.}

\begin{document}

\title{Mass, radius, and composition of the transiting planet \cnce\,:\\
using interferometry and correlations}

\correspondingauthor{Aur\'elien Crida}
\email{crida@oca.eu}

\author{Aur\'elien Crida}
\affiliation{Universit\'e C\^ote d'Azur / Observatoire de la C\^ote d'Azur --- Lagrange (UMR\,7293),\\
Boulevard de l'Observatoire, CS 34229, 06300 Nice, \textsc{France}}
\affiliation{Institut Universitaire de France, 103 Boulevard Saint-Michel, 75005 Paris, \textsc{France}}

\author{Roxanne Ligi}
\affiliation{INAF-Osservatorio Astronomico di Brera, Via E. Bianchi 46, I-23807 Merate, Italy}

\author{Caroline Dorn}
\affiliation{University of Zurich, Institut of Computational Sciences,\\
University of Zurich, Winterthurerstrasse 190, CH-8057, Zurich, \textsc{Switzerland}}

\author{Yveline Lebreton}
\affiliation{LESIA, Observatoire de Paris, PSL Research University, CNRS, Universit\'e Pierre et Marie Curie, Universit\'e Paris Diderot, 92195 Meudon, \textsc{France}}
\affiliation{Institut de Physique de Rennes, Université de Rennes 1, CNRS UMR 6251, 35042 Rennes, \textsc{France}}



\begin{abstract}
The characterization of exoplanets relies on that of their host star.
However, stellar evolution models cannot always be used to derive the
mass and radius of individual stars, because many stellar internal
parameters are poorly constrained.  Here, we use the probability
density functions (PDFs) of directly measured parameters to derive the
joint PDF of the stellar and planetary mass and radius. Because
combining the density and radius of the star is our most reliable way
of determining its mass, we find that the stellar (respectively
planetary) mass and radius are strongly (respectively moderately)
correlated.  We then use a generalized Bayesian inference analysis to
characterize the possible interiors of \cnce. We quantify how our
ability to constrain the interior improves by accounting for
correlation. The information content of the mass-radius correlation is
also compared with refractory element abundance constraints. We
provide posterior distributions for all interior parameters of
interest.  Given all available data, we find that the radius of the
gaseous envelope is $0.08 \pm 0.05 R_p$.  A stronger correlation
between the planetary mass and radius (potentially provided by a
better estimate of the transit depth) would significantly improve
interior characterization and reduce drastically the uncertainty on
the gas envelope properties.
\end{abstract}



\section{Introduction}
\label{sec:intro}
Following the era of detection that started with \citet{Mayor1995},
the characterization of exoplanets is one of the great scientific
adventures of the early 21st century. Transiting planets are
particularly interesting because their radius can be determined from
the transit depth. On top of this, transmission spectroscopy can
provide insights on their gas layers, if any. The satellites
\textit{CoRoT} \citep{Baglin2003} and \textit{Kepler}
\citep{Borucki2010} have been dedicated to the study of stellar light
curves and the search for exoplanetary transits, with remarkable
success. The light curves are so fine that the transit depth can be
determined with amazing precision (less than 2\% in 125 cases
referenced on \texttt{exoplanets.org}). Follow-up with spectrographs
such as HARPS \citep{Mayor-etal-2003} then provides the amplitude of
the radial velocity signal, from which the planet-to-star mass ratio
can be deduced. Despite an inherent degeneracy, the ability to
characterize the interiors of exoplanets improves with higher
precision on mass and radius. To date, 2379 objects have both a mass
and a radius in the \texttt{exoplanets.org} database (which includes
unconfirmed candidates), but only 100 with a precision better than
$5\%$ for both quantities. High-precision data are the challenge of
the next decade. In many cases, the uncertainty on planetary
parameters is dominated by the uncertainties in mass and radius (which
are generally of several percent) of the host star. We will never know
a planet better than its host star. This is why the new missions
dedicated to the search for transiting planets --\,CHEOPS
\citep{Broeg2013}, TESS \citep{Ricker2014}, and PLATO
\citep{Rauer-etal-2014}\,-- now focus on bright stars, whose
properties can be more easily determined by ground-based
instruments. In particular, one of the most important parameters
needed to characterize exoplanets is the stellar radius \citep[see
  e.g.][]{Creevey2007}. If the star is brighter than $\sim 8$
mag, it can be obtained by interferometry
\citep[see][]{Mourard-etal-2009, Ligi2014,Ligi-etal-2015} with better
than $2\%$ precision \citep[e.g.][]{Kervella2004,
  Boyajian2012a,Boyajian2012b, Ligi-etal-2012,Ligi-etal-2016}.
 
One of the few bright stars hosting transiting planets known today is
\cnc\ (a.k.a. HIP\,43587, HD\,75732, $\rho$1 Cnc A). This star is the
main component of a wide binary system, and hosts a system of five
planets, detected with the radial velocity technique \citep[][and
  references therein]{Fischer-etal-2008}. One of them (\cnce, the
closest to the star) is transiting and has been detected independently
by \citet{Winn-etal-2011} and \citet{Demory2011}. As one of the first
transiting super-Earths, it has received a lot of attention, and many
studies have already attempted to determine its composition. Previous
studies employed infrared and optical observations of transits,
occultations, and phase curves
\citep{Demory-etal-2012,Demory-etal-2016,Angleo-Hu-2017}. The planet
is highly irradiated with an equilibrium temperature of about
$2000$~K. The phase curve analysis revealed a large day--night-side
temperature contrast ($\sim$ $1300$~K) and a shift of the hottest spot
to the east of the substellar point
\citep{Demory-etal-2016,Angleo-Hu-2017}. The implication for a
possible gas layer is an optically thick layer with inefficient heat
redistribution.  The presence of a hydrogen-rich layer is unlikely,
since it would not sustain stellar evaporation and in fact no extended
hydrogen atmosphere has been detected (\citealp{Ehrenreich-etal-2012};
but see \citealp{Tsiaras-etal-2016}). If a gas layer is present, it
would be of secondary (enriched) nature
\citep{Dorn-Heng-2017}. Furthermore, the study of \cnce's thermal
evolution and atmospheric evaporation by \citet{Lopez-2017} suggests
either a bare rocky planet or a water-rich interior. But a bare rocky
planet is disfavored by \citet{Angleo-Hu-2017} and
\citet{Dorn-etal-2017b}.  The composition of \cnce\ is a matter of
debate and a consistent explanation of all observations is yet to
come.

The most recent interferometric study of \cnc\ was performed by
\citet{Ligi-etal-2016}, who provide a determination of the stellar
angular diameter with $1.64\%$ precision, independent of any stellar
evolution model (although a limb darkening model was used). Their work
is consistent within $1\%$ with a previous angular diameter estimate by
\citet{vanBraun2011}. Since \cnc\ hosts a transiting exoplanet, the
density of the star was determined using the transit light curve
by \citet{Maxted-etal-2015_mass-age}, and thus, \citet{Ligi-etal-2016}
derived the stellar mass directly with $7\%$ uncertainty. It is
therefore timely to use these new data to constrain the internal
structure of the transiting planet.

In this paper, we present in sections~\ref{sec:star}
and~\ref{subsec:MRp} a general method to rigorously make use of all
available interferometric observations, reducing the uncertainty and
using the correlations between the various stellar parameters. As much
as possible, we use analytical derivations of the probability density
functions (PDFs) of the parameters of interest from those of the observed
quantities. We apply these numerically to the case of \cnc\ and its
transiting planet, and show that we can reduce the uncertainty on the
planetary density. In section~\ref{subsec:IC}, these new estimates of
the planetary mass and radius and their correlation are used to
determine the internal composition of \cnce, using the model of
\citet{Dorn-etal-2017a}. Compared to previous applications of the
model \citep{Dorn-etal-2017b}, we have a slightly different estimate
for the mass and radius of the planet, and we account for the
correlation between them as well as for asymmetric uncertainties. The
results are then compared to a scenario where the mass-radius
correlation is neglected, and to a scenario where constraints on
refractory element abundances are used. Thereby, we can quantify the
information content of the different data inputs on the planetary
interior. Eventually, we provide the most precise interior estimates while
rigorously accounting for data uncertainties. Section~\ref{sec:conclu}
is devoted to a summary and conclusion.

\section{Stellar parameters\,: A joint PDF}

\label{sec:star}

In this section, we focus on the parameters of the host star,
\cnc. The observational quantities are the transit lightcurve, the
angular diameter $\theta$, the spectral energy distribution from which
we derive the bolometric flux $F_{\rm bol}$, and the parallax
$\px$. We combine them to retrieve the parameters of interest
(luminosity $L_\star$, effective temperature $\Tf$, mass $M_\star$,
radius $R_\star$). More specifically, we provide analytically the
joint PDF of these parameters from that of the observable
quantities. A joint PDF shows the correlations\,; from the way the
parameters are derived, correlations are strong and inevitable, and
provide valuable information, as will be illustrated in this
paper. Also, multiplying by a prior may lead to non-Gaussian final
distributions.

\subsection{PDF of the stellar mass and radius from observations only\,:
 A Bayesian approach}

Before determining mass and radius of \cnc, we first evaluate prior
knowledge on stellar parameters that will help to improve the
interpretation of observational data. More specifically, we look for
possibilities of excluding sets of parameters that would correspond to
the less populated regions of the Hertzsprung--Russell (hereafter H-R)
diagram.  We take a Bayesian approach in order to estimate $L_\star$
and $\Tf$. In essence, this approach accounts for both the probability
distribution of $L_\star$ and $\Tf$ for the star \cnc\ as deduced from
observations of the star, and the prior distribution of $L_\star$ and
$\Tf$ for stars in general as derived from the H-R diagram. In the
following, we discuss the approach in more detail and explain how it
can affect the estimate of the stellar radius.

\subsubsection{Probability density function of the stellar radius}

The stellar radius $R_\star$ is the product of the angular radius
($\theta/2$, in radian) with the distance $d$, which is proportional
to the inverse of the parallax $\px$\,:
\begin{equation}
R_\star=\frac{\theta d}{2} = R_0\theta/\px
\end{equation}
where $R_0$ is a length. If $\theta$ is given in milliarcseconds (mas) and $\px$ in
arcseconds (as), $R_0=\frac{1 \rm pc}{2\,\mr} = 0.1075\,R_\odot$ (where
$\mr$ is the number of mas in one radian).

Therefore, the PDF of $R_\star$, $f_{R_\star}$, can be expressed as a
function of those of $\theta$ and $\px$ (respectively denoted $f_\theta$
and $f_{\px}$) as (see Appendix A):
\begin{eqnarray}
\label{eq:fR_p}
f_{R_\star}(R) & = &
\frac{1}{R_0}\int_0^\infty p\, f_{\px}(p)f_\theta\left(\frac{p\,R}{R_0}\right)\ \dd p\\
 & = &
\frac{R_0}{R^2}\int_0^\infty t\, f_{\px}\left(\frac{R_0\,t}{R}\right)f_\theta(t)\ \dd t \ .
\label{eq:fR_t}
\end{eqnarray}
Note that if $f_{\px}$ and $f_\theta$ are Gaussian functions, then
$f_{R_\star}$ is also a Gaussian of mean $R_0\theta_0/{\px}_{0}$ and
variance the sum of the variances of $\theta$ and $\px$, but this
expression is more general. It gives directly the PDF of $R_\star$
as a function of the observables.

The stellar radius is also linked to the stellar luminosity and
effective temperature by
\begin{equation}
R_\star = \sqrt{\frac{L_\star}{4\pi\ssb}}\Tf^2
\end{equation}
where $\ssb$ is the Stefan--Boltzmann constant. From this, the PDF of
$R_\star$ can also be expressed as a function of $f_{\rm HR}$, the joint
PDF of $L_\star$ and $\Tf$ (see Appendix A):
\begin{eqnarray}
\label{eq:fR_T}
\hspace{-1cm}f_{R_\star}(R)
 & = & \frac{2}{R}\int_{t=0}^\infty L_{(R,t)}\, f_{\rm HR}(L_{(R,t)},t)\ \dd t\\
 & = & \frac{1}{2R}\int_{l=0}^\infty T_{(R,l)}\, f_{\rm HR}(l,T_{(R,l)})\ \dd l
\label{eq:fR_L}
\end{eqnarray}
where $L_{(R,t)} = 4\pi R^2\ssb t^4$ and $T_{(R,l)} =
\left(\frac{l}{4\pi R^2\ssb}\right)^{1/4}$. With these expressions, we
can make use of a prior in the $L_\star$--$\Tf$ plane to infer the PDF
of $R_\star$.

\subsubsection{Likelihood and prior in the H-R diagram}

\begin{figure}
\includegraphics[width=\linewidth]{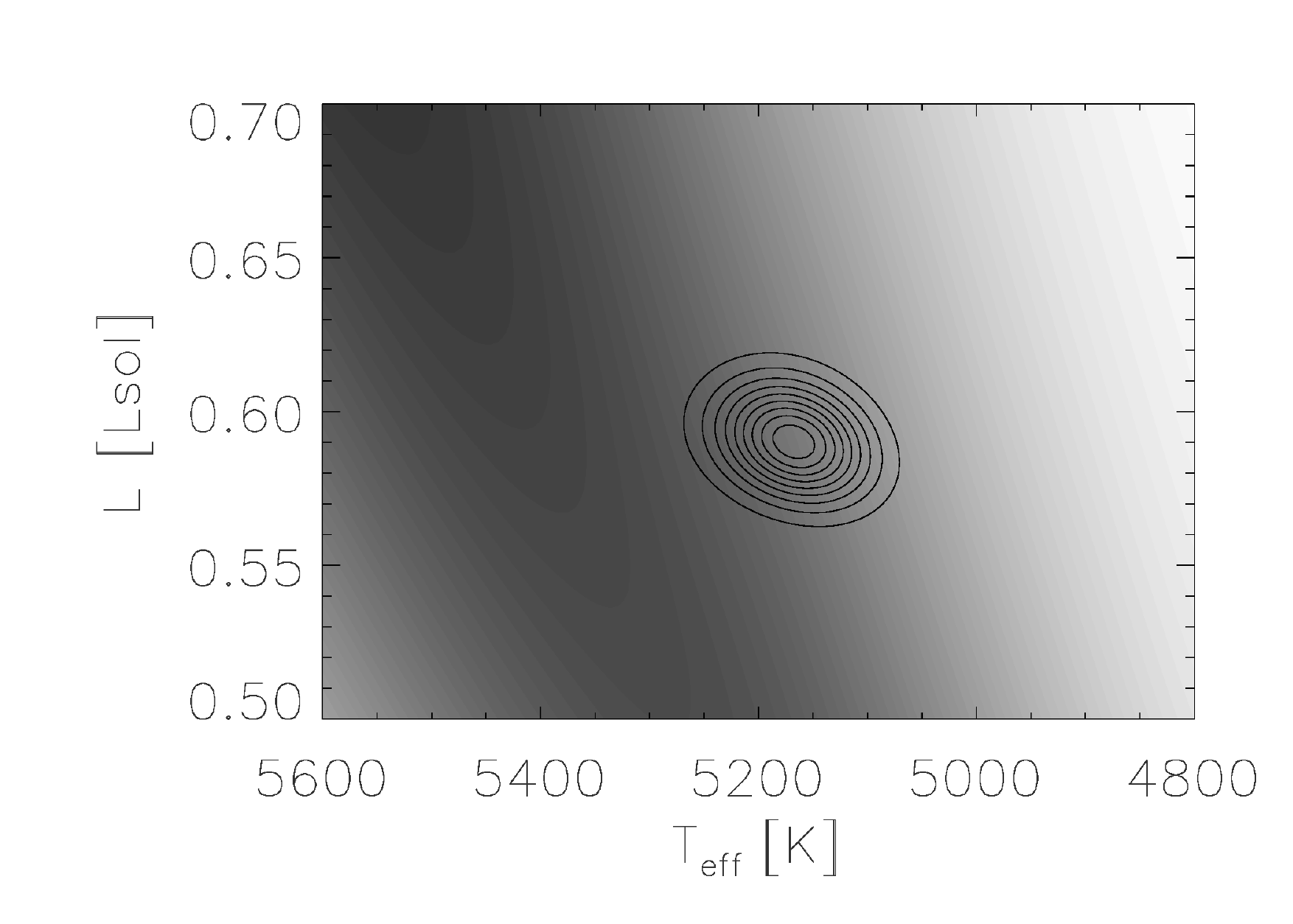}
\caption{\textbf{Contour lines\,:} likelihood $\Ld_{\rm HR}$ of the
  luminosity and effective temperature of \cnc\ as given by
  \myeqref{eq:Ld_LT} based on observations by
  \citet{Ligi-etal-2016}\,; nine contours separate 10 equal-sized
  intervals between 0 and the maximum of the
  likelihood. \textbf{Background grayscale\,:} density of stars in the
  \textit{Hipparcos} catalog in this region\,; in this box, the minimum and
  maximum of $f_{\rm Hip}^0$ are respectively 23 (light gray) and 488
  (dark).}
\label{fig:LT}
\end{figure}

\paragraph{Likelihood} The formulas linking $F_{\rm bol}$,
  $\theta$ and $\px$ to $L_\star$ and $\Tf$ are specified in
\citet{Ligi-etal-2016}, where the distributions of these two
parameters were computed separately using a standard propagation of
errors. Here, we derive analytically the joint likelihood of any pair
$(L_\star,\Tf)$ in the H-R plane, given the observational data
$f_{F_{\rm bol}}$, $f_{\px}$, $f_\theta$ (see Appendix B)\,:
\begin{equation}
\Ld_{\rm HR}(L_\star,\Tf) = \frac{{4\,\rm pc}\,\sqrt{\pi}\mr}{\Tf^{\,3}\sqrt{\ssb L_\star^{\,3}}}\times\int_0^{+\infty}\dd t
\label{eq:Ld_LT}
\end{equation}
$$\times t\times f_{F_{\rm bol}}(t)\times f_{\px}\left(\sqrt{\frac{4\pi t}{L_\star}}\right)\times f_\theta\left(\sqrt{\frac{4\,t}{\ssb\,\Tf^{\,4}}}\right)\ .$$
Taking $f_{F_{\rm bol}}$, $f_{\px}$ and $f_\theta$ as Gaussian
distributions of means and standard deviations as given in
\citet{Ligi-etal-2016}, we integrate numerically the expression above
and obtain for \cnc\ the contour lines shown in \figref{fig:LT}. They are
spread along a diagonal direction (along $L_\star\propto \Tf^4$, that
is equal radius lines) because both are increasing functions of
$F_{\rm bol}$ \citep[see also the Appendix of][]{Ligi-etal-2016}. From
\myeqref{eq:Ld_LT}, one can see that if the parallax and the angular
diameter were perfectly known (that is, if $f_{\px}$ and $f_\theta$ were
Dirac functions), $\Ld_{\rm HR}(L_\star,\Tf)$ would be non zero only on
the parametric curve $L_\star(t)=4\pi t/\px^2$, $\Tf(t) =
(4t/\ssb\theta^2)^{1/4}$. In this case, the correlation would be
$1$. This curve corresponds to varying $F_{\rm bol}$ while keeping the
stellar radius and distance fixed. The uncertainty on the stellar
radius and distance smears the PDF around this curve. Hence, the
better $\px$ and $\theta$ are constrained compared to $F_{\rm bol}$,
the more $L_\star$ and $\Tf$ are correlated. Here, the coefficient of
correlation of $L_\star$ and $\Tf$ is $0.23$\,.

\paragraph{Prior} \cnc\ is part of the \textit{Hipparcos} catalog, in
which the density of stars in the $(L_\star-\Tf)$ plane is not
uniform. Hence, one can estimate \emph{a priori} regions in the H-R
diagram where \cnc\ has more chances to be found, and regions where it
should not. This is a \emph{prior} PDF in the $(L_\star-\Tf)$
plane. To build this prior, we have downloaded the \textit{Hipparcos} catalog
\texttt{hip2.dat}\footnote{\tt
  ftp://cdsarc.u-strasbg.fr/pub/cats/I/239/hip\_main.dat.gz}, and
computed $L_\star$ and $\Tf$ for each star within $68.5$~pc from the
Sun as explained in detail in Appendix C.

In \figref{fig:LT}, the background grayscale maps $f_{\rm Hip}^0$, the
number density of stars in the \textit{Hipparcos} catalog (light for low
density, dark for high density, linear arbitrary scale). The main
sequence goes down steeply from the top left corner. Inside the
largest ellipse shown, the ratio of the maximum to minimum is $1.7$\,;
and within half the maximum of the likelihood, it is
$1.33$\,. The star \cnc\ appears to be in the vicinity of the main
sequence.

Eventually, the joint PDF of $L_\star$ and $\Tf$ is
\begin{equation}
f_{\rm HR}(L_\star,\Tf) = \Ld_{\rm HR}(L_\star,\Tf) \times f_{\rm Hip}^0(L_\star,\Tf)
\label{eq:fHR}
\end{equation}
It should be noted that $L_\star$ is so well constrained by the
observations that the multiplication by the prior has almost no effect
on the PDF of $L_\star$\,: we estimate $0.591\pm 0.013\,L_\odot$ from
$\Ld_{\rm HR}$ and from $f_{\rm HR}$ as well. As for the temperature, while
the expected value of $\Tf$ from $\Ld_{\rm HR}$ is $5169\,K$ with a
standard deviation of $46\,K$, the $\Tf$ found from $f_{\rm HR}$ is\,:
$5174 \pm 46\,K$.

The Kullback--Leibler divergence $$\mathcal{D}=\iint
f_{\rm HR}\ \ln\left(\frac{f_{\rm HR}}{f_{\rm Hip}^{\,0}}\right)\ \dd
L_\star\,\dd \Tf$$ is positive ($\sim 2.1$ when $ L_\star$ and $\Tf$
are integrated over a range of plus or minus $6\sigma$ around the
mean), and only $3\%$ smaller than using a uniform prior. The data are
very informative, and we are not dominated by the prior.

\subsubsection{Final Joint PDF of the Mass and Radius Using the Density}
\label{subsec:MR}

Using Equations~(\ref{eq:fR_p}) and (\ref{eq:fR_t}) gives $R_{\rm \cnc} = 0.960 \pm
0.0181\, R_\odot = (668.3 \pm 12.6)10^6$~m, $f_{R_\star}$ being a
Gaussian, as in \citet{Ligi-etal-2016}.

\begin{figure}
\includegraphics[width=\linewidth]{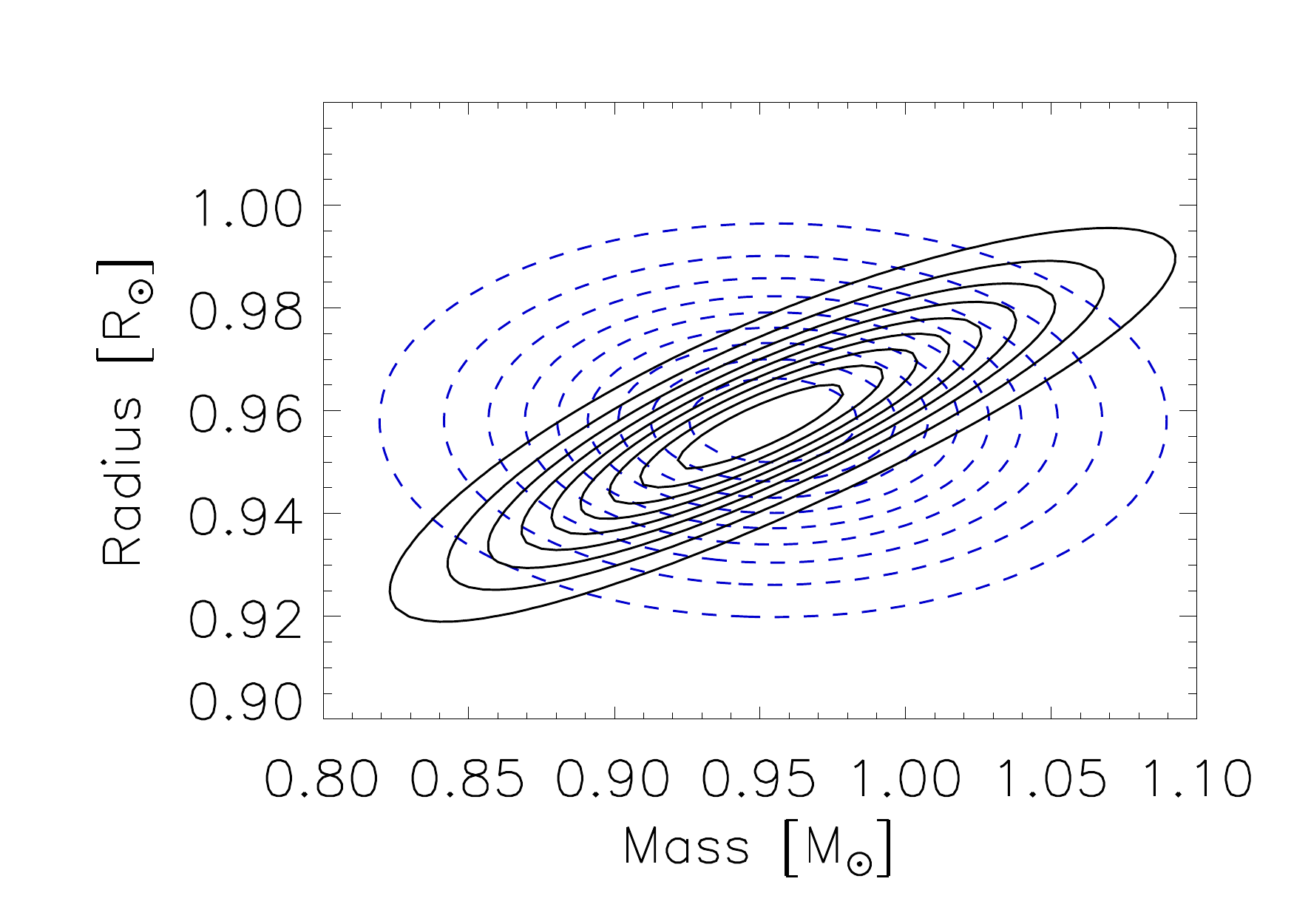}
\includegraphics[width=0.49\linewidth]{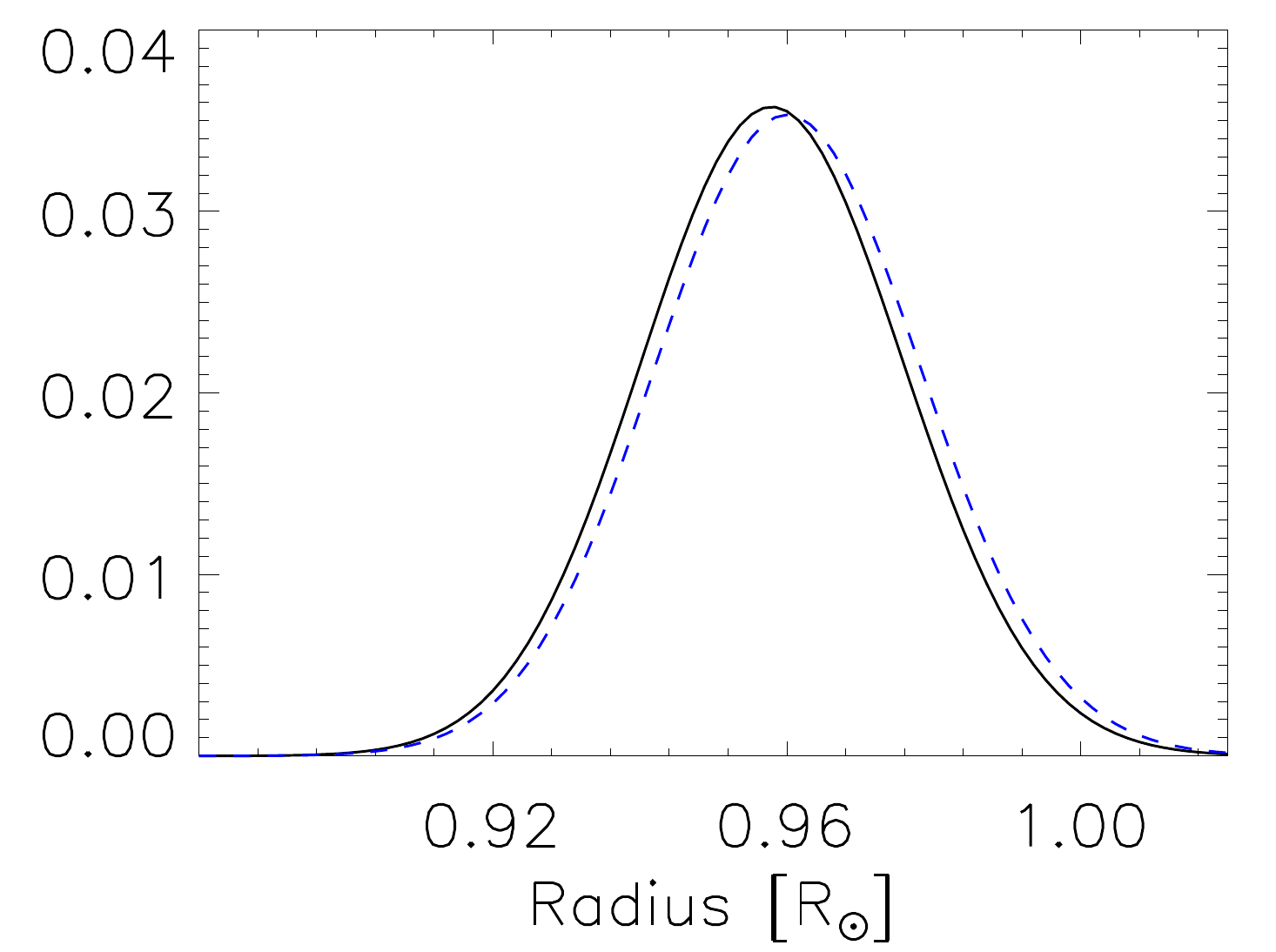}
\includegraphics[width=0.49\linewidth]{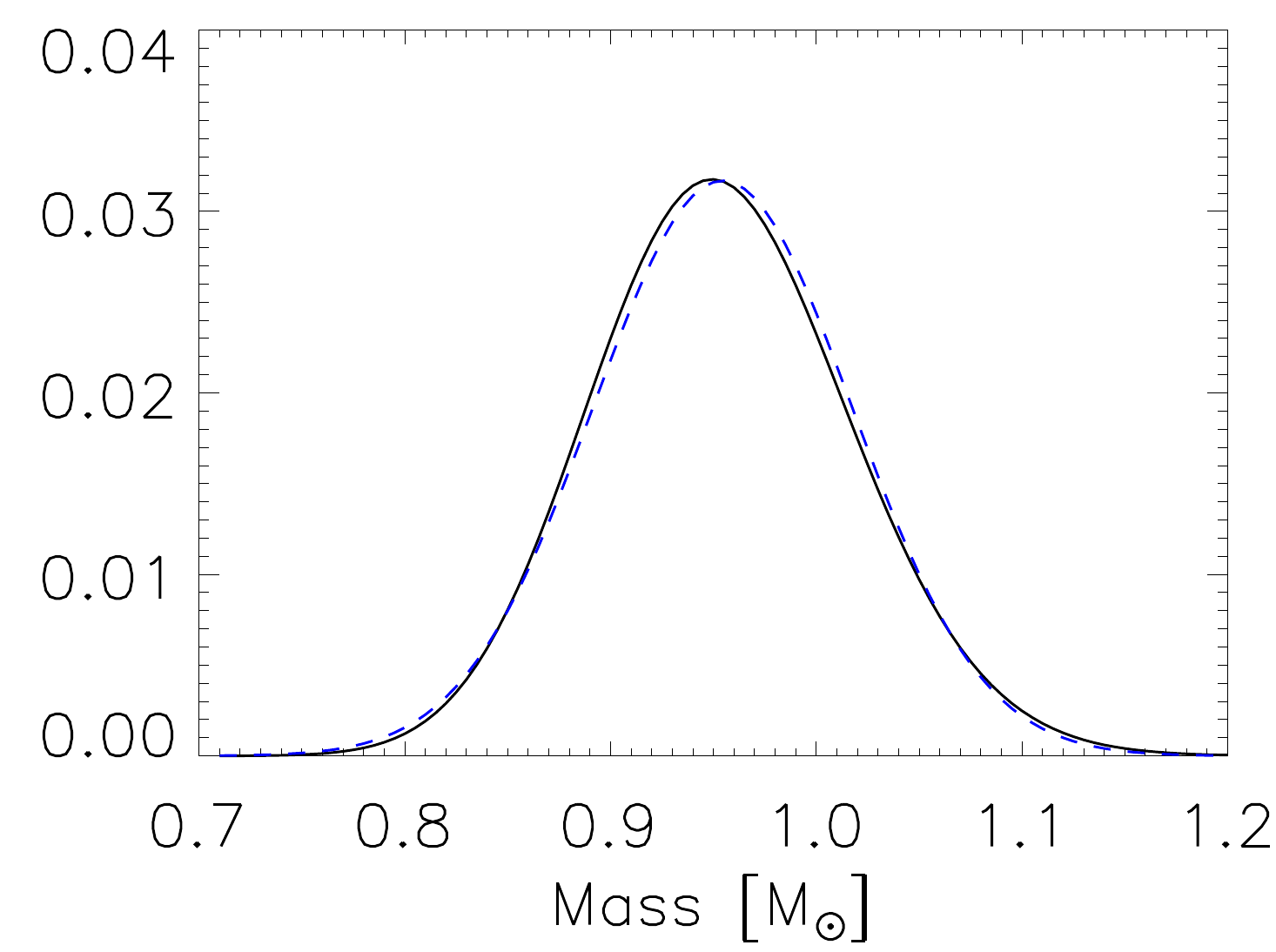}
\caption{\textbf{Top\,:} joint probability density function of the
  mass and radius of the star \cnc. The nine plain thick contour lines
  separate 10 equal-sized intervals between 0 and the maximum of
  Eq.~(\ref{eq:MRstar_anal}). The dashed blue contour lines show the
  same for the case where one mistakenly considers $M_\star$ and
  $R_\star$ as independent. \textbf{Bottom\,:} marginal PDFs of
  $R_\star$ and $M_\star$ (plain lines); the dashed blue line is the
  Gaussian obtained without the use of the prior in the case of
  $R_\star$, and is a Gaussian curve of the same mean and standard
  deviation, for comparison, in the case of $M_\star$.}
\label{fig:MRstar}
\end{figure}

In Appendix A.3, we show that using Equations~(\ref{eq:fR_T}) and (\ref{eq:fR_L})
with $f_{\rm HR}$ given by \myeqref{eq:Ld_LT} is exactly equivalent to
directly using Equations~(\ref{eq:fR_p}) and (\ref{eq:fR_t}). No information is
lost, and no uncertainty is added by moving to the H-R plane. Hence,
using Equations~(\ref{eq:fR_T}) and (\ref{eq:fR_L}) with $f_{\rm HR}$ given by
Equation~(\ref{eq:fHR}) shows only the effect of the prior. Integrating this
numerically, we find $R_{\rm \cnc} = 0.958\pm 0.0178\,R_\odot$. These
two PDFs of $R_\star$ are shown in the bottom left panel of
\figref{fig:MRstar}.

\citet{Maxted-etal-2015_mass-age} provide the density of \cnc\,:
$\rho_\star = 1.084 \pm 0.038\ \rho_\odot$. Indeed, a careful
analysis of the light curve, combining the transit period and the
transit duration directly yields the stellar density
$\rho_\star$ \citep{Seager-MallenOrnelas-2003}. Then, the joint
likelihood of $M_\star$ and $R_\star$ can be expressed analytically\,:
\begin{equation}
\Ld_{MR\star} (M,R) = 
\frac{3}{4\pi R^3}\times f_{R_\star}(R)\times f_{\rho_\star}\left(\frac{3M}{4\pi R^3}\right)
\label{eq:MRstar_anal}
\end{equation}
(see Appendix D). Using $f_{R_\star}$ given by
Eqs.~(\ref{eq:fR_p}-\ref{eq:fR_t}), the result is $M_{\rm \cnc} =
0.961 \pm 0.064\ M_\odot$, with a correlation coefficient with $R_{\rm
  \cnc}$ of $0.85$. The level curves of this distribution are shown in
\figref{fig:MRstar} as the tilted solid ellipses. Using the prior in
the H-R diagram, one gets $M_{\rm \cnc} = 0.954 \pm 0.063\ M_\odot$,
with a correlation coefficient with $R_{\rm \cnc}$ of $0.85$.

Our results are summarized and compared to the ones of
\citet{Ligi-etal-2016} in Table~\ref{tab:params}. We find that the
prior from the \textit{Hipparcos} catalog does not change significantly the
joint PDF of ($M_{\rm \cnc}$, $R_{\rm \cnc}$). The interferometric
observations are precise enough to constrain the stellar
parameters. In what follows, we thus use the analytical expressions
Equations.~(\ref{eq:fR_p}), (\ref{eq:fR_t}), and
(\ref{eq:MRstar_anal}).

If correlation is neglected and $M_\star$ and $R_\star$ are directly
taken with their uncertainties as independent variables, their joint
PDF becomes a 2D Gaussian distribution represented by the dashed
ellipses with horizontal and vertical axes in
\figref{fig:MRstar}. In doing so, one would have correct marginal
distributions (they are close to Gaussian). But one would mistakenly
consider likely combinations of $M_\star$ and $R_\star$ that can
actually be excluded by the constraint on $\rho_\star$. Obviously,
taking the correlation into account reduces the area to explore in the
mass-radius parameter plane, and should help constrain the
structure and composition of the transiting planet, as we will see in
the next section.

\subsection{About stellar models}

$L_\star$ and $\Tf$ of \cnc\ being known, one could fit them with
stellar evolution models to infer the corresponding mass, age, and
other parameters like the radius. Stellar models are a precious tool
to estimate stellar parameters that are not measurable, provided
observational constraints are tight enough. Nonetheless, this method
should be used with care, for the following reasons.
\begin{enumerate}
\item Degeneracy\,: low-mass stars gather on the main sequence where
  they slowly increase their luminosity and temperature for billions
  of years, inducing a huge mass--age degeneracy. In the case of \cnc,
  which is close to the main sequence, the degeneracy is
  between a pre- and a post-main sequence phase \citep[coined
    ``young'' and ``old'' solutions in][]{Ligi-etal-2016} ; the
  detection of lithium in its atmosphere
  \citep{Hinkel-etal-2014,Ramirez-etal-2014} advocates for the young
  solution.
\item Internal source of error\,: models are more or less sensitive to
  many parameters that are not always well constrained, such as the
  metallicity (with very different values provided in the literature
  for \cnc), the initial helium abundance, the rotation rate, and the
  choice of input physics. Assuming a default value of these
  parameters may lead to inaccuracy in the final result (see below).
\item External source of error\,: different models available in the
  literature can give different results, in part because of the two
  difficulties mentioned above \citep[see][]{Lebreton2012}.
\end{enumerate}
In fact, using the CES2MO pipeline\footnote{The CES2MO tool is a
  stellar model optimization pipeline. It has been described in
  \citet{Lebreton-Goupil-2014} and is based on the Cesam2k stellar
  evolution code \citep{Morel-Lebreton-2008}.} and our value for
$L_\star$ and $\Tf$, we find, for the young solution of \cnc, masses
ranging from $0.950\pm 0.015$ to $0.989\pm 0.020\,M_\odot$, depending
on the choice on the internal parameters (mostly the stellar
metallicity). This highlights the difficulty of using stellar models
to derive accurately the mass and radius of an individual star with
reliable uncertainties. Of course, accuracy is difficult to assess\,;
however, the variability of estimates yields a proxy for it. Here, the
different values from stellar models are only in rough agreement with
one another, so it would be inappropriate to just pick one, neglecting
the uncertainty on the parameters of the model.

Note that the mass range we find using the Bayesian approach above
encompasses the various stellar models mentioned here for the young
solution \citep[see also][]{Ligi-etal-2016}.  Although the
interferometric radius disagrees with the radius found by
asteroseismology for some stars (which opens the question of possible
bias for one of these methods), it overcomes assumptions that are
otherwise introduced by the use of stellar models. Hence, reassured by
the agreement with stellar models, in the following we adopt the
estimate of the mass and radius for \cnc\ given in
sect.~\ref{subsec:MR}. We stress that our error bar is larger than the
brutal use of a single stellar model could provide, but we think it is
the best possibility so far for \cnc.

%
%
%

\begin{table}
\caption{Properties of the star \cnc\ and of its transiting exoplanet \cnce.}
\label{tab:params}
\centering
\begin{tabular}{lrrl}
\hline
\hline
\multicolumn{4}{c}{Coordinates} \\
\hline
\multicolumn{2}{l}{R.A. (J2000)}     & \multicolumn{2}{c}{08h\ 52min\,35.81093s} \\ 
\multicolumn{2}{l}{Decl. (J2000)}    & \multicolumn{2}{c}{+28$^{\circ}$\,19'\,50.9511''} \\
\multicolumn{2}{l}{Parallax [mas]} & \multicolumn{2}{c}{81.03 $\pm$ 0.75} \\ 
\multicolumn{2}{l}{Distance [pc]}  & \multicolumn{2}{c}{12.34 $\pm$ 0.11} \\ 
\hline
\hline
\multicolumn{4}{c}{Stellar parameters} \\
                & Ligi+(2016) & This work & (corr.)\\
\hline
$M_\star$ [$M_\odot$] & 0.960 $\pm$ 0.067 & 0.954 $\pm$ 0.063 & \multirow{2}{*}{$0.85$}\\
$R_\star$ [$R_\odot$] & 0.96 $\pm$ 0.02   & 0.958 $\pm$ 0.018 & \\
$\rho_\star$ [$\rho_\odot$] & \multicolumn{2}{c}{1.084 $\pm$ 0.038} & \\
$L_\star$ [$L_\odot$] & 0.589 $\pm$ 0.014 & 0.591$\pm$ 0.013 & \multirow{2}{*}{$0.23$}\\
$\Tf$ [K]    &  5165 $\pm$ 46    & 5174 $\pm$ 46    & \\
\hline
\hline
\multicolumn{4}{c}{Planetary parameters} \\
                & Ligi+(2016) & This work & (corr.)\\
\hline
$M_{\rm p}$ [$M_\oplus$] & 8.631 $\pm$ 0.495 & 8.703 $\pm$ 0.482 & \multirow{2}{*}{$0.30$} \\
$R_{\rm p}$ [$R_\oplus$] & 2.031$^{+0.091}_{-0.088}$& 2.023 $\pm$ 0.088 & \\
$\rho_{\rm p}$ [$\rho_\oplus$] & $ 1.03\pm 0.14$ & $1.06\pm 0.13$ & \\
\hline
\end{tabular}
\end{table}

\section{Planetary parameters and composition}

\label{sec:planet}

In this section, we apply the previous results on the host star to the
transiting planet \cnce. This planet has attracted a lot of attention
already, being one of the first discovered transiting super-Earth, as
explained in Section~\ref{sec:intro}. It is therefore an excellent
case to test the power of our method.

\subsection{Likelihood and joint PDF}

\label{subsec:MRp}

From the PDF of the mass and radius of the star, we deduce that of the
planet analytically. For any $M_p$, $M_\star$, one can define the
associated semi-amplitude of the radial velocity signal $K$, following
a classical formula resulting from Kepler's law:
$K(M_p,M_\star)=\frac{M_p}{M_\star^{2/3}}\left(\frac{2\pi
  G}{P}\right)^{1/3}$ (where $P$ is the orbital period, and we have
assumed that the eccentricity is zero\footnote{The eccentricity of
  \cnce\ is $0.028$ in \texttt{exoplanet.eu}, which makes the
  assumption $e\approx 0$ reasonable.}). Similarly, for a pair $R_p$,
$R_\star$, the corresponding transit depth is
$\TD(R_p,R_\star)=(R_p/R_\star)^2$. Therefore, the PDF associated to
any fixed planetary mass and radius is
\begin{eqnarray*}
f_p(M_p,R_p) & \propto & \iint \exp\left(-\frac12\left(\frac{K(M_p,M_\star)-K_e}{\sigma_K}\right)^2\right) \\
 & & \ \times \exp\left(-\frac12\left(\frac{\TD(M_p,M_\star)-\TD_e}{\sigma_{\TD}}\right)^2\right)\\
 & & \ \times \Ld_{MR\star}(M_\star,R_\star)\  \dd M_\star\,\dd R_\star
\label{eq:PDF_MR_planet}
\end{eqnarray*}
where the observed transit depth associated to \cnce\ is
$\TD_e\pm\sigma_{\TD}=(3.72\pm 0.30) 10^{-4}$
\citep{Dragomir-etal-2014}, and the amplitude of the signal in radial
velocity is $K_e\pm\sigma_K=6.30\pm 0.21$~m\,s$^{-1}$ \citep{Endl-etal-2012}.


\begin{figure}
\includegraphics[width=\linewidth]{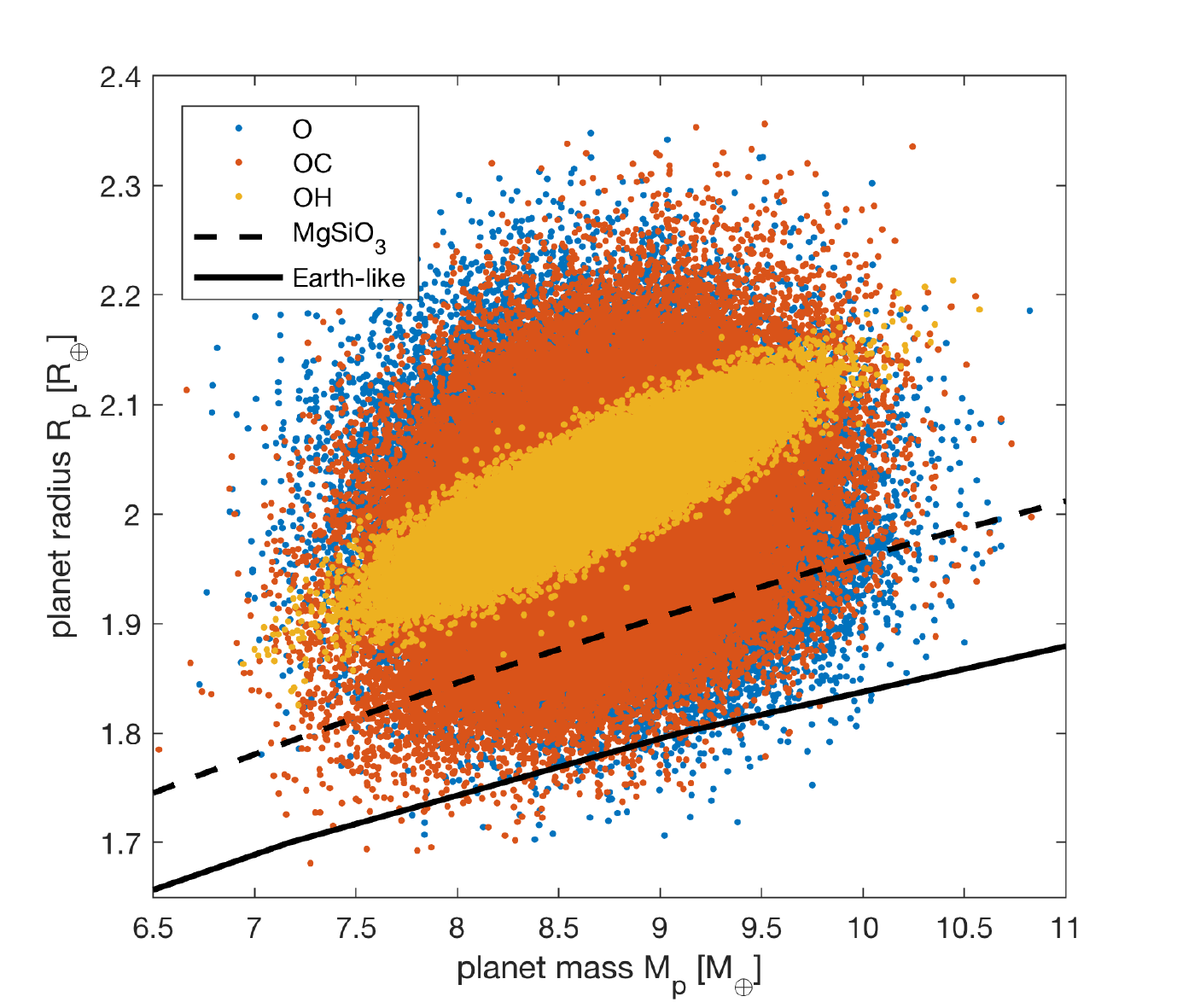}
\caption{Mass and radius data samples for \texttt{O}, \texttt{OC},
  and the \texttt{OH} that mostly differ in terms of the correlation
  between mass and radius. In comparison, two idealized mass-radius
  relationships for pure MgSiO$_3$ and Earth-like interiors are
  plotted. MgSiO$_3$ represents the least dense end-member of purely
  rocky interiors. Therefore, purely rocky interiors cannot be exluded
  in cases of \texttt{O} and \texttt{OC}, whereas in the case of the
  hypothetical high correlation (\texttt{OH}), the interior must be
  rich in volatiles. See the text for details.}
\label{fig:MRSAMP}
\end{figure}

This expression has been integrated numerically\,; we find\,:
\begin{eqnarray}
\label{eq:Mp}
M_p & = & 8.703 \pm 0.482\ M_\oplus\\
\label{eq:Rp}
R_p & = & 2.023 \pm 0.088\ R_\oplus
\end{eqnarray}
with a correlation of $c=0.30$.\\


The cloud of red dots labeled \texttt{OC} in \figref{fig:MRSAMP} shows
a Monte Carlo realization of this PDF. The correlation is visible, as
the cloud is elongated in a direction parallel to isodensity lines. An
Earth-like composition is almost excluded, while a pure rocky interior
appears possible. The blue dots in \figref{fig:MRSAMP} correspond to
the case where $\Ld_{MR\star}(M_\star,R_\star)$ would be replaced in
the expression of $f_p(M_p,R_p)$ by a PDF of $M_\star,R_\star$ that
would neglect their correlation (shown as short dashed lines in
\figref{fig:MRstar}). In this case, an Earth-like composition could be
excluded with less confidence.

It is particularly interesting to consider the correlation in order to
estimate the density of the planet. From our joint PDF, we find
$\rho_p = 5846\pm 740$~kg\,m$^{-3} = 1.06\pm
\,0.13\rho_\oplus$~\footnote{A careful reader may notice that
  $8.703/2.023^3=1.051$, not $1.06$. Because $<R_p^{\,3}>\neq<R_p>^3$,
  the expected value of $\rho_p$ is not given by $<M_p>/<R_p>^3$.}. A
standard propagation of errors assuming $M_p$ and $R_p$ indepenent
would give $\rho_p = 5797 \pm 819$~kg.m$^{-3}$. We get a $10\%$
smaller uncertainty on the density of \cnce\ taking the correlation
into account. The limiting factor here is the uncertainty on $\TD_e$,
which is mainly responsible for the correlation between mass and
radius to be much smaller for the planet ($0.30$) than for the host
star ($0.86$). Indeed, the $8\%$ uncertainty on $\TD_e$ translates
into $4\%$ in the radius ratio, while the stellar radius is determined
to within $2\%$. More precise observations of the transit would be
very useful in this particular case and would allow us to increase
significantly the gain on the density precision. On the other hand,
the $3\%$ uncertainty on $K_e$ is smaller than that on $M_\star$ (and
even on $M_\star^{2/3}$) so, to gain precision in the planetary mass,
one should aim at gaining precision on the stellar mass. In the
particular case of \cnc, the best way to do so would be to better
constrain its density by obtaining a finer light curve\footnote{Note
  added after publication\,: this has been done just a few weeks after
  the publication of this article by \citet{Bourrier+18}. See
  \citet{Crida+2018b} for an update of this paper. For reference, we
  eventually find $M_p = 8.59 \pm 0.43\ M_\oplus$, $R_p = 1.947 \pm
  0.038\ R_\oplus$, with a correlation of $c=0.54$.}.

In the next subsection, we use this joint PDF to characterize the
interior of \cnce, including a test scenario where $\TD_e$ and $K_e$
would be known with negligible uncertainty, which is shown in
\figref{fig:MRSAMP} as the pale dots labeled \texttt{OH}\,; in this
case, one recovers the $0.85$ correlation associated with the
distribution of the stellar mass and radius.


\subsection{Structure and Composition}

\label{subsec:IC}

\subsubsection{Method}
The estimates of planetary mass and radius are subsequently used to
characterize the interior of \cnce. To do so, we use the
generalized Bayesian inference analysis of \citet{Dorn-etal-2017a}
that employs a Markov chain Monte Carlo (MCMC) method. This method
allows us to rigorously quantify the degeneracy of the following
interior parameters for a general planet interior:
\begin{enumerate}\itemsep 0pt
\item core: core size (\rc),
\item mantle: mantle composition (mass ratios $\fesima$,
  $\mgsima$) and size of rocky interior (\rsolid),
\item gas: intrinsic luminosity (\Lenv), gas mass (\menv), and
  metallicity (\Zenv).
\end{enumerate}
In this study, the planetary interior is assumed to be composed of a
pure iron core, a silicate mantle comprising the oxides
Na$_2$O--CaO--FeO--MgO--Al$_2$O$_3$--SiO$_2$, and a gas layer of H,
He, C, and O. Unlike \citet{Dorn-etal-2017b}, we have assumed no
additional water layer. For the highly irradiated planet \cnce,
any water layer would be in a vapour or super-critical state.

The prior distributions of the interior parameters are listed in
Table~\ref{tableprior}. The priors are chosen conservatively. The
cubic uniform priors on \rc and \rsolid reflect equal weighing of
masses for both core and mantle. Prior bounds on $\fesima$ and
$\mgsima$ are determined by the host star's photospheric abundance
proxies, whenever abundance constraints are considered. Otherwise,
$\fesima$ and $\mgsima$ are chosen such that the iron oxide can range
from $0\%$ to $70\%$ in weight while the magnesium and silicate oxides
can range from $0\%$ to $100\%$ (all oxides summing up to $100\%$ of
course). Since iron is distributed between core and mantle, $\fesi$
only sets an upper bound on $\fesima$. A log-uniform prior is set for
\menv and \Lenv.

In general, the data that we consider as input to the interior
characterization are:
\begin{enumerate}\itemsep 0pt
\item Original data (\texttt{O}), that comprises the planetary mass
  and radius given by Eqs.~(\ref{eq:Mp}) and (\ref{eq:Rp}), the
  orbital radius, and the stellar irradiation (namely, stellar
  effective temperature $T_{\rm eff} = 5174 K$ and stellar radius
  $R_\star = 0.961 R_{\odot}$).
\item Correlation (\texttt{C})  between mass and radius: $c$ =0.30,
\item Abundances (\texttt{A}), that comprise bulk abundance constraints on
  $\fesi$ and $\mgsi$, and minor elements Na, Ca, Al. From the stellar
  ratios that can be measured in the stellar photosphere, one gets:
  $\fesi$ = 1.86 $\pm$ 1.49, $\mgsi$ = 0.93 $\pm$ 0.77, m$_{\rm CaO}$
  = 0.013 wt\%, m$_{\rm Al_2O_3}$ = 0.062 wt\%, m$_{\rm Na_2O}$ =
  0.024 wt\% \citep{Dorn-etal-2017b}.
\end{enumerate}
We consider different scenarios labeled \texttt{O}, \texttt{OC},
\texttt{OA}, and \texttt{OCA} where the letters correspond to the set
of data taken into account. For example, for the data scenario
\texttt{O}, we consider planetary mass and radius as well as other
data, but we neglect mass-radius correlation and abundance
constraints.

The structural model for the interior uses self-consistent
thermodynamics for core, mantle, and to some extent also the gas
layer.  For the core density profile, we use the equation of state
(EoS) fit of iron in the hexagonal close-packed structure provided by
\citet{bouchet} on {\it ab initio} molecular dynamics simulations.
For the silicate mantle, we compute equilibrium mineralogy and density
as a function of pressure, temperature, and bulk composition by
minimizing the Gibbs free energy \citep{connolly09}.  We assume an
adiabatic temperature profile within core and mantle.

For the gas layer, we solve the equations of hydrostatic equilibrium,
mass conservation, and energy transport. For the EoS of elemental
compositions of H, He, C, and O, we employ the Chemical
Equilibrium with Applications package \citep{CEA}, which performs
chemical equilibrium calculations for an arbitrary gaseous mixture,
including dissociation and ionization and assuming ideal gas
behavior. The metallicity \Zenv is the mass fraction of C and O in the
gas layer, which can range from 0 to 1.  For the gas layer, we assume
an irradiated layer on top of a convection-dominated layer, for which
we assume a semi-gray, analytic, global temperature averaged profile
\citep{Guillot-2010, Heng2014}. The boundary between the irradiated
layer and the underlying layer is defined where the optical depth in
visible wavelength is $100 / \sqrt{3}$ \citep{JIN2014}. Within the
convection-dominated layer, the usual Schwarzschild criterion is used
to determine where in the layer convection or radiation is more
efficient.  The planet radius is defined where the chord optical depth
becomes 0.56 \citep{Lecavelier08}.
We refer the reader to model I in \citet{Dorn-etal-2017a} for more
details on both the inference analysis and the structural model.

\begin{table}
\caption{Prior Ranges.  \label{tableprior}}
\centering
\begin{tabular}{lll}
\hline\noalign{\smallskip}
Parameter & Prior Range & Distribution  \\
\noalign{\smallskip}
\hline\noalign{\smallskip}
$r_{\rm core}$         & (0.01  -- 1) $r_{\rm core+mantle}$ &uniform in $r_{\rm core}^3$\\
$\fesima$           & 0 -- $\fesistar$&uniform\\
$\mgsima$         & $\mgsistar$ &Gaussian\\
\rsolid   & (0.01 -- 1) $R$& uniform in $r_{\rm core+mantle}^3$\\
\menv            & 0 -- $m_{\rm env, max}$  & uniform in log-scale\\
\Lenv                & $10^{18} - 10^{23}$ erg\,s$^{-1}$ & uniform in log-scale\\
\Zenv                & 0 -- 1&uniform\\
\hline
\end{tabular} 
\end{table}

\begin{table*}[ht]
\caption{Interior Parameter Estimates for Different Scenarios.
  \label{tableresults}}
\begin{center}
\begin{tabular}{l|rrrrr}
\hline\noalign{\smallskip}
Interior Parameter & \hspace{1.2cm}\texttt{O}\hspace{0.5cm} \  & \ \hspace{1.2cm}\texttt{OC}\hspace{0.5cm} \ & \ \hspace{1.2cm}\textbf{\texttt{OCA}}\hspace{0.5cm} \ & \ \hspace{1.2cm}\texttt{OA}\hspace{0.5cm} \ & \ \hspace{1.2cm}\texttt{OH}\hspace{0.5cm} \ \\
\noalign{\smallskip}
\hline\noalign{\smallskip}
log$_{10}$({\menv/M$_p$}) & $-4.75_{-1.74}^{+2.03}$ & $-4.86_{-1.71}^{+2.03}$ & {\boldmath$-5.07_{-1.61}^{+2.14}$} & $-5.32_{-1.87}^{+2.14}$ & $-4.49_{-1.49}^{+1.97}$ \\
\Zenv & $0.55_{-0.29}^{+0.23}$ & $0.55_{-0.29}^{+0.23}$ & {\boldmath$0.58_{-0.30}^{+0.22}$} & $0.57_{-0.30}^{+0.23}$ & $0.55_{-0.30}^{+0.21}$ \\
log$_{10}$(\Lenv)& $21.46_{-2.11}^{+2.12}$ & $21.51_{-2.11}^{+2.08}$ & {\boldmath$21.49_{-2.14}^{+2.13}$} & $21.48_{-2.14}^{+2.14}$ & $21.48_{-2.15}^{+2.13}$\\
$r_{\rm gas}$& $0.09_{-0.05}^{+0.06}$ & $0.09_{-0.05}^{+0.05}$ & {\boldmath$0.08_{-0.05}^{+0.05}$} & $0.08_{-0.06}^{+0.05}$ & $0.10_{-0.03}^{+0.05}$ \\
\rsolid /R$_p$& $0.91_{-0.06}^{+0.05}$ & $0.91_{-0.05}^{+0.05}$ & {\boldmath$0.92_{-0.05}^{+0.05}$} & $0.92_{-0.05}^{+0.06}$ & $0.90_{-0.05}^{+0.03}$ \\
$r_{\rm core}$/\rsolid & $0.41_{-0.14}^{+0.13}$ & $0.40_{-0.13}^{+0.13}$ & {\boldmath$0.36_{-0.12}^{+0.10}$} & $0.35_{-0.11}^{+0.10}$ & $0.39_{-0.12}^{+0.13}$ \\
$\fesima$& $6.47_{-4.36}^{+7.25}$ & $6.69_{-4.54}^{+7.83}$ & {\boldmath$1.31_{-0.85}^{+1.19}$} & $1.37_{-0.88}^{+1.19}$ & $6.84_{-4.68}^{+8.52}$ \\
$\mgsima$& $6.83_{-4.16}^{+5.80}$ & $6.97_{-4.15}^{+5.74}$ & {\boldmath$1.03_{-0.57}^{+0.66}$} & $1.04_{-0.58}^{+0.66}$ & $7.14_{-4.20}^{+5.83}$ \\
\hline
\end{tabular}\vspace{6pt}

{\small Note. Uncertainties of 1-$\sigma$ are listed.\\ We use the
\texttt{OCA} scenario (in bold) for the final interpretation of
possible interiors of 55Cnc e.}
\end{center}
\end{table*}

\subsubsection{Results}

\begin{figure*}
\includegraphics[width=\textwidth]{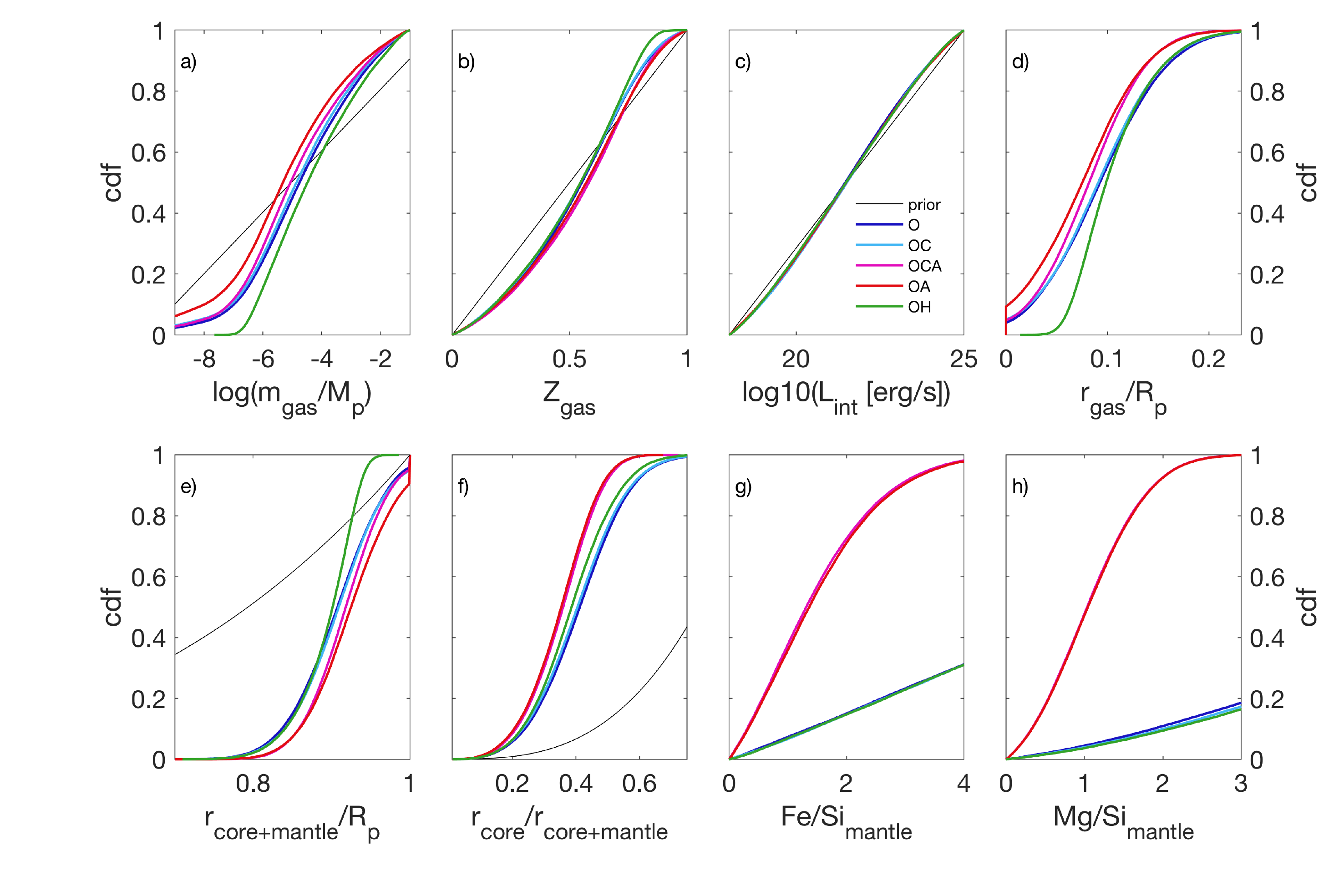}
\caption{Sampled one-dimensional marginal posterior for interior
  parameters: (a) gas mass fraction \menv, (b) gas metallicity \Zenv ,
  (c) intrinsic luminosity \Lenv, (d) gas radius fraction, (e) size of
  rocky interior \rsolid/$R_{\rm p}$, (f) relative core size $r_{\rm
    core}$/\rsolid, (g), (h) mantle composition in terms of mass ratios
  $\fesima$ and $\mgsima$. The prior distributions are shown in black.
  For (g), (h) the priors vary between the data scenarios (\texttt{O},
  \texttt{OC}, \texttt{OH} versus \texttt{OCA}, \texttt{OA}) and are
  not shown.}
\label{fig:IC1}
\end{figure*}

We investigate the information content of the different data scenarios
labeled \texttt{O}, \texttt{OC}, \texttt{OA}, and
\texttt{OCA}. For each scenario, we have used the generalized MCMC
method to calculate a large number of sampled models ($\sim 10^6$)
that represent the posterior distribution of possible interior
models. The resulting posterior distributions are shown in
\figref{fig:IC1}, which displays cumulative distribution functions
(cdf). The thin black line is the initial (prior) distribution. The
colored lines correspond to the different data scenarios. They
indicate how the ability to estimate interiors changes by considering
different data. A summary of interior parameter estimates is stated in
Table \ref{tableresults}.

In the first scenario (\texttt{O}), the uncorrelated planetary mass
and radius given in Table~\ref{tab:params} are considered, as well as
the orbital radius and stellar luminosity. These data help to
constrain the mass and radius fraction of the gas layer, the size of
the rocky interior and the core, while intrinsic luminosity, gas
metallicity, and mantle composition are poorly constrained. In the
second scenario (\texttt{OC}), we add the correlation coefficient of
$M_p$ and $R_p$. Since this correlation is low ($c = 0.3$, see also
\figref{fig:MRSAMP}), differences in our ability to constrain the
interior are marginal\,: uncertainty ranges for \rsolid , $r_{\rm
  core}$, \menv, and $r_{\rm gas}$ reduce by $\sim
1\%$. 

In the \texttt{OA} scenario, we add constraints on refractory element
ratios compared to the scenario \texttt{O} with uncorrelated mass and
radius. The abundance constraints significantly improve estimates on
the mantle composition (by $\sim$ 85\%) and the core size (by $\sim$
20\%). Thereby the density of the rocky interior is better constrained
which also affects the estimates of \rsolid , \menv, and $r_{\rm gas}$
by a few percent. The information value of abundance constraints is
discussed by \citet{Dorn-etal-2015} in detail.

If abundance constraints are considered, the effect of adding the
mass-radius correlation is more pronounced. This can be seen by
comparing scenario \texttt{OA} with \texttt{OCA}, in which the
latter scenario accounts for both the correlation and the abundance
constraints. The additional correlation mostly improves \rsolid ,
\menv, and $r_{\rm gas}$. The 10th percentiles (and 90th percentiles)
of the gas radius fraction (and the rocky radius fraction) change by
2\% compared to the planet radius.

To study the importance of the mass-radius correlation, we add a
hypothetical scenario (\texttt{OH}), in which the uncertainty on the
transit depth $TD_e$ and radial velocity signal $K_e$ are assumed
negligible, such that the correlation between the planetary mass and
radius is equal to that between the stellar mass and radius with $c =
0.869$. Note that neglecting the uncertainty on the planet-to-star
radius and mass ratios also leads to reducing significantly the
uncertainties on $M_p$ and more importantly $R_p$\,: we get $R_p=2.025
\pm 0.042\ R_\oplus$ (where the slight but negligible difference in
the expected value with the previous case is due to the non-use of the
\textit{Hipparcos} prior here).  For \texttt{OH}, we generally find
that interior estimates significantly improve compared to
\texttt{OCA}. This is true for \rsolid, \menv, and $r_{\rm gas}$. In
this scenario, we can exclude the possibility of a purely rocky
interior and find gas layers with radius fractions larger than 0.05
and mass fractions larger than $10^{-7}$. This (hypothetical) case
illustrates the high value in both a high radius precision and
mass-radius-correlation for interior characterization.

\begin{figure*}
\vspace{-10pt}
\includegraphics[width=\textwidth]{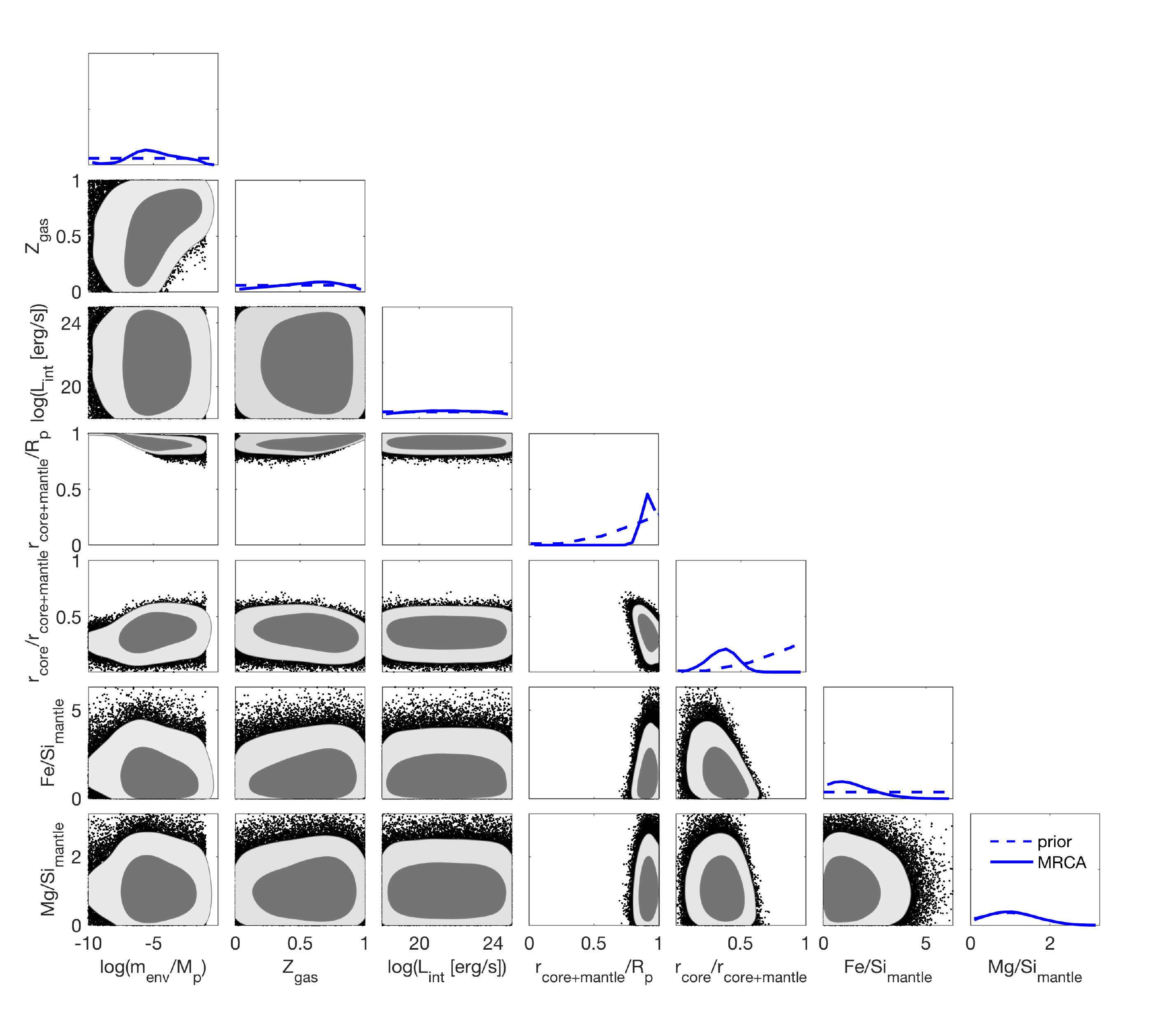}
\caption{Sampled two and one-dimensional marginal posterior for all
  interior parameters and the \texttt{OCA} data scenario. Grey
  shaded 2-D areas represent 1-$\sigma$ and 2-$\sigma$ distributions
  of marginalized posteriors. Prior distributions are shown in dashed
  blue for the one-dimensional marginal posteriors. }
\label{fig:IC2}
\end{figure*}

The \texttt{OCA} scenario represents our most complete dataset given
the considered interferometric data. Figure \ref{fig:IC2} shows the
posterior distribution of the \texttt{OCA} scenario in more
detail. The one-dimensional posterior functions illustrate that only
some interior parameters can be constrained by data, since prior and
posterior distributions significantly differ: gas mass fraction \menv,
\rsolid, $r_{\rm core}$, and $\fesima$. The gas layer properties of
metallicity and intrinsic luminosity are very degenerate and the data
considered here do not allow to constrain them.  We find that the gas
layer has a radius fraction of $r_{\rm gas} = 0.08\pm 0.05\ R_p$ and
a mass fraction about 10 times larger than for Earth, although with
large uncertainty (see Table~\ref{tableresults}). The gas metallicity
is weakly constrained; however, low metallicities are less likely i.e.,
there is an 80\% chance that the metallicity is larger than 0.3
(while assuming a uniform prior on \Zenv). The size of the rocky
interior is estimated to be \rsolid = $0.92 \pm 0.05\ R_p$ with a
core of size $r_{\rm core} = 0.36^{+0.10}_{-0.12}$ \rsolid.

Between the scenarios \texttt{O}, \texttt{OC}, \texttt{OH} on one hand
and \texttt{OCA}, \texttt{OA} on the other, there is a large
difference in the predicted range of mantle compositions. For the
former, the ratios of $\fesima$ and $\mgsima$ are large, albeit with
huge uncertainties, while for the latter these ratios are
significantly better constrained, due to the used abundance
constraints ($\fesi$ and $\mgsi$). Note that a larger $\fesima$
induces a denser mantle, hence a thicker gas layer. These differences
illustrate the high information value of abundance constraints for
which the stellar composition may be used as a proxy
\citep{Dorn-etal-2015} in order to reduce the otherwise high
degeneracy. Only mass and radius (\texttt{O}, \texttt{OC},
\texttt{OH}) allow for a large range of possibly unrealistic mantle
compositions that are very different from Earth-like mantle
compositions (Mg/Si$\sim 1$ and Fe/Si $< 1$).

\subsubsection{Discussion}

An alternative interior scenario could include C-rich
compositions. Such interiors are indeed possible, and have been
proposed in the past \citep[e.g.][]{Madhusudhan-etal-2012}. This was
motivated by a high C/O ratio estimate for the star
\citep[$1.12\pm0.19$,][]{DelgadoMena-etal-2010}, but this ratio has
been later corrected down to $0.78\pm0.08$ \citep{Teske-etal-2013},
making C-rich interior models less timely for \cnce. Although
  \citet{Moriatry-etal-2014} argue that a sequential condensation
  during the whole life of an evolving proto-planetary disk can favor
  the formation of C-rich planetesimals, they find that the
  planetesimals expected to form around \cnc\ should have C/O$<$1,
  even assuming C/O=1 for this system (their figure~1). In addition,
C-rich interiors are poorly understood. Some exotic models exist that
account SiC, C, and Fe layers, but neglect major rock-forming elements
(e.g. Mg, O) \citep{Kuchner-Seager2005,Bond-etal-2010}. In order to
make meaningful predictions on C-rich interior structures, a better
understanding of carbon-bearing compounds, their phase diagrams, phase
equilibria, and EoSs are required \citep[e.g., Miozzi et
  al. (in review)][]{Nisr-et-al-2017, Wilson-and-Militzer-2014}.

For reference, assuming a C-rich interior for the planet could lead to
a larger \rsolid because SiC can be less dense than silicates (in its
zinc-blende (B3) form), hence to a thinner gas layer\,; but again
these models suffer from large uncertainties. In particular,
\citet{Daviau-Lee-2017b} show that B3 SiC decomposes into Si and C
(diamond) above roughly 2000~K, which is likely to apply to \cnce's
mantle. Also, \citet{Daviau-Lee-2017a} find that B3 SiC transitions to
a rocksalt (B1) form at high pressures, which has a density very close
to that of MgSiO$_3$. This would make an SiC planet undistinguishable
from a silicate one from the mass radius relation only. It would also
conveniently make our conclusions on the size of the mantle
independent of whether it is made of silicates or of B1 SiC.

\section{Conclusions}

\label{sec:conclu}

In this paper, we have characterized the possible interiors of
\cnce\ starting with a rigorous investigation of the observations of its
host star. Compared to previous work, we have adopted a more analytical
approach, which allows us to use a prior in the H-R diagram and to obtain
semi-analytically the joint PDF of the mass and radius of the star,
then of the planet. We have estimated the uncertainties on these
parameters carefully, taking inherent correlations into
account. Besides the particular case of \cnce, our analysis helps to
demonstrate the information value of different data types besides mass
and radius: mass-radius-correlation and refractory element abundances.

We provide an analytical expression for the joint likelihood of the
stellar luminosity and temperature directly from the observables. This
formula allows us to skip a Monte Carlo analysis. In the case of \cnc,
we find that the stellar parameters are well enough constrained by
interferometry with respect to our prior based on the
\textit{Hipparcos} catalog, which does not bring much significant
information. The distribution of the stellar mass and radius is also
derived analytically; they are very strongly correlated, thanks to the
constraint on the stellar density. Compared to stellar evolution
models, our stellar parameters are in good agreement, with an
uncertainty encompassing the various outcomes of different models. We
conclude that stellar evolution models are good in general, but should
be used with great care for the case of individual stars: they
provide appealing small uncertainty, but their accuracy is very
sensitive to many parameters. The method we developed here seems to be
a more reliable way of estimating stellar and thus planetary mass and
radius, because it is based on direct measurements, and in particular
that of the stellar radius (unfortunately not always available). Of
course, if the age of the planet is needed (e.g., in the case of gas
giant planets that contract as they evolve), stellar models would be a
necessary step to infer it, via the dating of the host star.

Using the planetary mass and radius that we derived, we inferred the
internal structure of the planet \cnce, using the model developed by
\citet{Dorn-etal-2017a}. Our results show that the data on mass and
radius, taken independently, allow to estimate the internal structure
of the planet to some degree. Improved estimates can be obtained by
accounting for (1) possible correlation of mass and radius or (2)
abundance constraints that were discussed in previous studies.
In the case of \cnce, the $0.3$ correlation is too small to have
significant influence on interior estimates. In any case, there is
a well-known inherent degeneracy such that a large number of interiors
can fit even infinitely precise mass and radius. Assuming that the
planet's $\fesima$ and $\mgsima$ are similar to the star's helps
contrain the internal structure of the planet much better, in
particular the size of the core and the mantle composition, which is
only poorly constrained by the mass-radius correlation.

We find that there is a low chance, of 5\%, that the interior is purely
rocky. The gas layer thickness is estimated to be 8\% ($\pm$ 5\%) of
the total radius. We stress that a more precise estimate of the
transit depth would allow us to increase significantly the mass-radius
correlation of the planet, and thus to reduce significantly the
uncertainty on the thickness and mass of the gaseous layer and the
rocky interior, as well as on the core size\footnote{The reader is referred to \citet{Crida+2018b} for an update of this work using new, better data.}.

\acknowledgments We warmly thank Diana Valencia for interesting
discussions on the internal composition of planets, Florentin Millour
for explanations concerning the \textit{Hipparcos} catalog, Georges
Kordopatis, Orlagh Creevey and Mathias Schultheis for insights on
stellar models and populations.\\ R.L. is funded by the European
Union's Horizon 2020 research and innovation programme under the Marie
Sk\l odowska-Curie grant agreement n. 664931.\\ C.D. is funded by the
Swiss National Science Foundation under the Ambizione grant
PZ00P2\_174028.



\appendix

\section{Probability Density Function of $R_\star$}

Let us deote $f_X$ as the PDF of $X$ and $F_X$ as its cumulative distribution function. 

\subsection{From the observations of the angular diameter and the parallax}
\label{subsec:Rfromthetapx}

The stellar radius $R_\star$ is the product of the angular radius
($\theta/2$) with the distance $d$, and the distance is proportional
to the inverse of the parallax $\px$. Thus, one can write
\begin{equation}
R_\star=\frac{\theta d}{2} = R_0\theta/\px
\end{equation}
where $R_0$ is a length, equal to $\frac{1 \rm pc}{2\,\mr} =
0.1075\,R_\odot$ if $\theta$ is in milliarcseconds (mas) and $\px$ in arcseconds (as).

As a consequence, $R_\star$ is lower than $R$ if and only if $\theta$
is lower than $\px(\frac{R}{R_0})$, whatever the value of $\px$. Thus,
the probability that $R_\star<R$ reads:
\begin{equation}
F_{R_\star}(R) = \int_0^\infty f_{\px}(p)\left[\int_0^{\frac{pR}{R_0}}f_\theta(t)\,\dd t\right]\ \dd p
\end{equation}
From this, one deduces the PDF of $R_\star$ as follows\,:
\begin{eqnarray*}
f_{R_\star}(R) & = & F_{R_\star}'(R)\\
 & = & \int_0^\infty f_{\px}(p)\frac{\partial}{\partial R}\left[\int_0^{\frac{pR}{R_0}}f_\theta(t)\,\dd t\right]\ \dd p\\
 & = & \int_0^\infty f_{\px}(p)\left(\frac{p}{R_0}\right)f_\theta\left(\frac{p\,R}{R_0}\right)\ \dd p\\
 & = & \frac{1}{R_0}\int_0^\infty p\, f_{\px}(p)f_\theta\left(\frac{p\,R}{R_0}\right)\ \dd p
\end{eqnarray*}

A change of variable ($t=p\,R/R_0$) gives the equivalent expression
used in the main text:
\begin{equation}
f_{R_\star}(R)  =  \frac{R_0}{R^2}\int_0^\infty t\, f_{\px}\left(\frac{R_0\,t}{R}\right)f_\theta(t)\ \dd t\ .
\label{eq:fR_thetapx}
\end{equation}

\subsection{From the joint PDF of ($L_\star,\Tf$)}
\label{RfromLT}
The stellar luminosity and effective temperature are connected through
the stellar radius as: $L_\star=4\pi R_\star^2\ssb\Tf^4$. Therefore,
$R_\star<R$ is equivalent to $L_\star<4\pi R^2\ssb\Tf^4$. Hence, with
$f_{\rm HR}$ the joint PDF of $L_\star$ and $\Tf$:
\begin{eqnarray*}
F_{R_\star}(R) & = & \iint_{\{l<4\pi R^2\ssb t^4\}} f_{\rm HR}(l,t)\ \dd l\ \dd t\\
 & = & \int_{t=0}^\infty \left[\int_0^{4\pi R^2\ssb t^4} f_{\rm HR}(l,t)\ \dd l\right]\, \dd t
\end{eqnarray*}
Again, derivation with resect to $R$ gives the PDF of $R_\star$:

\begin{eqnarray}
\hspace{-1cm}f_{R_\star}(R) & = & \int_{t=0}^\infty (8\pi R\ssb t^4)\,f_{\rm HR}(4\pi R^2\ssb t^4,t)\ \dd t\\
 & = & \frac{2}{R}\int_{t=0}^\infty L_{(R,t)}\, f_{\rm HR}(L_{(R,t)},t)\ \dd t
\end{eqnarray}
where $L_{(R,t)} = 4\pi R^2\ssb t^4$.\\
Noting $T_{(R,l)} = \left(\frac{l}{4\pi R^2\ssb}\right)^{1/4}$, and making the change of variable $l=L_{(R,t)}$ leads to the equivalent expression:
\begin{equation}
f_{R_\star}(R) =  \frac{1}{2R}\int_{l=0}^\infty T_{(R,l)}\, f_{\rm HR}(l,T_{(R,l)})\ \dd l
\label{eq:fR_LT}
\end{equation}

\subsection{Equivalence of the two methods}

Below, we show that \myeqref{eq:fR_LT} is exactly equivalent to
\myeqref{eq:fR_thetapx} if $f_{\rm HR}$ is taken as $\Ld_{\rm HR}$ derived
from $f_{F_{\rm bol}}$, $f_{\px}$, and $f_\theta$ in
Appendix~\ref{app:PDF_LT} (see \myeqref{eq:Ld_LT}\,). This means that
using \myeqref{eq:fR_LT}, one does not lose any information compared to
directly using $f_\theta$ and $f_{\px}$ with \myeqref{eq:fR_thetapx}\,:

\begin{eqnarray*}
f(R) & = & \frac{2}{R}\int_{t=0}^\infty L_{(R,t)} \Ld_{\rm HR}(L_{(R,t)},t)\, \dd t\\
 & = & \frac{2}{R}\int_{t=0}^\infty L_{(R,t)} \frac{{4 \rm pc}\sqrt{\pi/\ssb}\mr}{L_{(R,t)}^{3/2}\,t^3}
\times\left[\int_{\tau=0}^\infty \tau\, f_{F_{\rm bol}}(\tau)f_\theta\left(\mr\sqrt{\frac{4\tau}{\ssb t^4}}\right)f_{\px\!}\left(\sqrt{\frac{4\pi \tau}{L_{(R,t)}}}{1\rm pc}\right) \dd\tau \right]\, \dd t\\
 & = & \frac{4\,{\rm pc}\ \mr}{R^2}\int_{t=0}^\infty\frac{\dd t}{\ssb\,t^5} \left[\int_{u=0}^\infty \frac{\ssb t^4}{4\,\mr^2}u^2
f_{F_{\rm bol}}\left(\frac{\ssb t^4 u^2}{4\,\mr^2}\right) f_\theta(u) f_{\px\!}\left(\sqrt{\frac{\ssb\,t^4}{L_{(R,t)}}}\frac{u({1\rm pc})}{\mr}\right) \frac{\ssb t^4 u\,\dd u}{2\,\mr^2}\right]\\
 & = & \frac{{1\,\rm pc}}{2R^2\mr}\iint\dd u\ \dd t\ \ssb\,t^3\,u^3 
f_{F_{\rm bol}}\left(\frac{\ssb t^4 u^2}{4}\right) f_\theta(u) f_{\px\!}\left(\frac{u({1\rm pc})}{2R\mr}\right)\\
 & = & \frac{R_0}{R^2}\int_{u=0}^\infty \dd u\ f_\theta(u) f_{\px\!}\left(\frac{uR_0}{R}\right) \, u
\int_{t=0}^\infty \ssb t^3 u^2 f_{F_{\rm bol}}\left(\frac{\ssb t^4 u^2}{4}\right)  \dd t\\
 & = & \frac{R_0}{2R^2}\int_{u=0}^\infty \dd u\ f_\theta(u) f_{\px\!}\left(\frac{uR_0}{R}\right) \, u\,\underbrace{\int_{\phi=0}^\infty f_{F_{\rm bol}}\left(\phi\right)  \dd \phi}_{1}
\end{eqnarray*}

Hence, one can apply the prior $f_{\rm Hip}^0$ to the PDF of $R_\star$
by simply calculating
\begin{equation}
f_{R_\star}(R) =  \frac{1}{2R} \int_0^\infty L_{(R,t)}
   \Ld_{\rm HR}(L_{(R,t)},t)f_{\rm Hip}^0(L_{(R,t)},t)\, \dd t\ .
\label{eq:fR_prior}
\end{equation}

\section{Likelihood of $L_\star$ and $\Tf$, Given Obseravtions}
\label{app:PDF_LT}

Here, we want to derive analytically the likelihood of a pair of
luminosity and effective temperature against the observations of the
angular diameter, parallax, and bolometric flux. The PDFs of the
observables are denoted respectively $f_\theta$, $f_{\px}$ and
$f_{F_{\rm bol}}$. The likelihood in the H-R plane is denoted
$\Ld_{\rm HR}$.

 Be $H=\{L<a;T<b\}$ a subset of the universe
$\Omega=\{L\in\R+;T\in\R+\}$. The probability of $H$ is naturally
$$\Prob(H)\equiv P(a,b)=\int_{u=0}^{u=a}\int_{v=0}^{v=b} \Ld_{\rm HR}(u,v)\,\dd v\,\dd u\ .$$
Hence 
\begin{equation}
\Ld_{\rm HR}(a,b)=\frac{\partial^2P(a,b)}{\partial a\  \partial b}\ .
\label{eq:fLTderiv}
\end{equation}

$L_\star$ and $T_{\rm eff}$ are given as functions of the observable
quantities by\,:
\begin{eqnarray}
\label{eq:L}
L & = & 4\pi\,F_{\rm bol}\,\left(\frac{1\,\rm pc}{\px\,\rm[as]}\right)^2\\
\label{eq:T}
T_{\rm eff} & = & \left(\frac{4}{\ssb}\right)^{\!1/4}F_{\rm bol}^{\ 1/4}\,(\theta\,{\rm [rad]})^{-1/2}\ ,
\end{eqnarray}
where $\ssb$ is the Stefan--Boltzmann constant. Thus, $H$ can also be
defined as:
$$\left\{F_{\rm bol}=t\in\R\,; \ \px{\rm [as]}>{1\,\rm
  pc}\sqrt{\frac{4\pi t}{a}}\,; \ \theta{\rm [mas]} >
\mr\sqrt{\frac{4\,t}{\ssb\,b^4}}\right\}$$
(where $\mr=2.06\cdot 10^8$ is the number of mas in 1~rad). From now
on, $\theta$ is implicitely given in mas, and $\px$ in as. The
probability of the event $H$ is given by
\begin{equation}
\Prob(H) = \int_0^{+\infty} f_{F_{\rm bol}}(t)\times \left[1-F_{\px}\left(\sqrt{\frac{4\pi t}{a}}{1\,\rm pc}\right)\right]\times\left[1-F_\theta\left(\mr\sqrt{\frac{4\,t}{\ssb\,b^4}}\right)\right]\ \dd t
\label{eq:PdeH}
\end{equation}
Using Eqs.~(\ref{eq:fLTderiv}) and (\ref{eq:PdeH}), one obtains
\begin{eqnarray*}
\Ld_{\rm HR}(a,b) & = & \frac{\partial^2}{\partial a\  \partial b}P(a,b)\\
 & = & \frac{\partial}{\partial a}\Bigg\{ \int_0^{+\infty} f_{F_{\rm bol}}(t)\times \left[1-F_{\px}\left(\sqrt{\frac{4\pi t}{a}}{1\,\rm pc}\right)\right]
\times \frac{\partial}{\partial b}\left(\left[1-F_\theta\left(\mr\sqrt{\frac{4\,t}{\ssb\,b^4}}\right)\right]\right)\ \dd t\Bigg\}\\
 & = & \frac{\partial}{\partial a}\Bigg\{ \int_0^{+\infty} f_{F_{\rm bol}}(t)\times \left[1-F_{\px}\left(\sqrt{\frac{4\pi t}{a}}{1\,\rm pc}\right)\right]
\times \left[-\frac{-2\mr}{b^3}\sqrt{\frac{4t}{\ssb}}f_\theta\left(\mr\sqrt{\frac{4\,t}{\ssb\,b^4}}\right)\right]\ \dd t\Bigg\}\\
 & = & \int_0^{+\infty} f_{F_{\rm bol}}(t)\times\frac{\partial}{\partial a}\left\{\left[1-F_{\px}\left(\sqrt{\frac{4\pi t}{a}}{1\,\rm pc}\right)\right]\right\}
\times \left[\frac{4\mr}{b^3}\sqrt{\frac{t}{\ssb}}f_\theta\left(\mr\sqrt{\frac{4\,t}{\ssb\,b^4}}\right)\right]\ \dd t\\
 & = & \int_0^{+\infty} f_{F_{\rm bol}}(t)\times\left\{\frac{1\,\rm pc}{2}\sqrt{\frac{4\pi t}{a^3}}f_{\px}\left(\sqrt{\frac{4\pi t}{a}}{1\,\rm pc}\right)\right\}
\times \left[\frac{4\mr}{b^3}\sqrt{\frac{t}{\ssb}}f_\theta\left(\mr\sqrt{\frac{4\,t}{\ssb\,b^4}}\right)\right]\ \dd t
\end{eqnarray*}
\begin{equation}
\Ld_{\rm HR}(a,b) = \frac{{4\,\rm pc}\sqrt{\pi}\mr}{b^3\sqrt{\ssb a^3}}\times \int_0^{+\infty} f_{F_{\rm bol}}(t)\,f_{\px}\left(\sqrt{\frac{4\pi t}{a}}\right)f_\theta\left(\sqrt{\frac{4\,t}{\ssb\,b^4}}\right)\ t\,\dd t
\end{equation}

\section{Density of stars in the H-R plane in the solar neighborhood from the \textit{Hipparcos} catalog}

From the \textit{Hipparcos} catalog, we compute the effective temperature and
luminosity of each star as follows.
\begin{itemize}
\item The effective temperature is a function of the $B-V$ color index
  (provided in the catalog) given by \citet{Flower-1996} and \citet[][Table
  2]{Torres-2010}.
\item The luminosity $L_\star$ is given by\,:
\begin{equation}
2.5\,\log(L/L_\odot) = 4.74 - \underbrace{H_p+BC-5\,\log(1/\px\ [{\rm as}])}_{M_{\rm bol}}\ ,
\label{eq:Lum_Hip}
\end{equation}
where $M_{\rm bol}$ is the absolute bolometric magnitude ($4.74$ being
the solar absolute bolometric magnitude adopted here), with $H_p$ the
\textit{Hipparcos} magnitude, $BC$ the bolometric correction, and $\px$ the
parallax. $H_p$ and $\px$ are in the catalog. For $BC$, we fit
linearly \citet{Cayrel-etal-1997} in the region of interest for us
($5000\,K<\Tf<5500\,K$) as\,: $BC = -2.44 + 0.0004\,\Tf\,.$ We have
checked that a more elaborate functional form of $BC$ has no
significant impact on the density of stars near \cnc.
\end{itemize}

Then, the density of stars next to the point $(L_0,T_0)$ is defined as
\begin{equation}
f_{\rm Hip}^0(L_0,T_0) = \sum_{{\px}_{,i}>14.6\,\rm mas} \exp \left\{ - \frac12
\left(\frac{\log(L_0)-\log(L_i)}{0.08}\right)^2 -
\frac12\left(\frac{T_0-T_i}{100\,K}\right)^2 \right\}
\label{eq:rho_Hip}
\end{equation}
where the widths of the Gaussian kernels in $L_\star$ and $T$ have been
chosen to obtain a smooth density function in the region next to
\cnc\ without losing information. The sum goes through all the stars
of the catalog with a parallax larger than $14.6$\,mas (while that of
\cnc\ is $81$\,mas). Indeed, brighter stars can be seen from larger
distances, and hence would be overrepresented in the catalog without
a distance limit. The \textit{Hipparcos} catalog is complete up to a magnitude
$H_p = 8.5$, and we want our sample to be complete up to
$\log(L/L_\odot)=0.1$ to cover well the \cnc\ region of the HR
diagram. The limit parallax then results from \myeqref{eq:Lum_Hip}.

\section{Calculation of the joint PDF of $M_\star$ and $R_\star$ from the PDFs of $R_\star$ and $\rho_\star$}

The subset $K=\{M_\star<a;R_\star<b\}$ of the $M_\star-R_\star$ space
is identical to $\{\rho_\star < \frac{3 a}{4\pi
  R_\star^{\,3}};R_\star<b\}$. Hence,
$\Prob(K)=\displaystyle\int_{R_\star=0}^{\ b}\int_{\rho_\star=0}^{\frac{3 a}{4\pi
    R_\star^{\,3}}}f_{R_\star}(R_\star)f_{\rho_\star}(\rho_\star)\ \dd\rho_\star\,\dd
R_\star $~. 

\begin{eqnarray*}
\Ld_{MR_\star}(a,b) & = & \frac{\partial^2 \Prob(K)}{\partial a\ \partial b}\\
 & = & \frac\partial{\partial b}\int_{R_\star=0}^b
       f_{R_\star}(R_\star)\,\frac\partial{\partial a}
       \int_{\rho_\star=0}^{\frac{3 a}{4\pi R_\star^{\,3}}}f_{\rho_\star}(\rho_\star)\ 
       \dd\rho_\star\,\dd R_\star\\
 & = & \frac\partial{\partial b}\int_{R_\star=0}^b
       f_{R_\star}(R_\star)\,\frac{3}{4\pi R_\star^{\,3}}\,f_{\rho_\star}
       \left(\frac{3a}{4\pi R_\star^{\,3}}\right)\ \dd R_\star\\
 & = & f_{R_\star}(b)\,\frac{3}{4\pi b^{\,3}}\,f_{\rho_\star}
       \left(\frac{3a}{4\pi b^{\,3}}\right)\\
\Ld_{MR_\star}(a,b) & = & \frac{3}{4\pi b^{\,3}}\,f_{\rho_\star}
       \left(\frac{3a}{4\pi b^{\,3}}\right)\,f_{R_\star}(b)
\end{eqnarray*}




\bibliographystyle{aasjournal}
\bibliography{./Cnc.bib}

\begin{thebibliography}{}
\expandafter\ifx\csname natexlab\endcsname\relax\def\natexlab#1{#1}\fi
\providecommand{\url}[1]{\href{#1}{#1}}
\providecommand{\dodoi}[1]{doi:~\href{http://doi.org/#1}{\nolinkurl{#1}}}
\providecommand{\doeprint}[1]{\href{http://ascl.net/#1}{\nolinkurl{http://ascl.net/#1}}}
\providecommand{\doarXiv}[1]{\href{https://arxiv.org/abs/#1}{\nolinkurl{https://arxiv.org/abs/#1}}}

\bibitem[{{Angelo} \& {Hu}(2017)}]{Angleo-Hu-2017}
{Angelo}, I., \& {Hu}, R. 2017, \aj, 154, 232, \dodoi{10.3847/1538-3881/aa9278}

\bibitem[{{Baglin}(2003)}]{Baglin2003}
{Baglin}, A. 2003, Advances in Space Research, 31, 345,
  \dodoi{10.1016/S0273-1177(02)00624-5}

\bibitem[{{Bond} {et~al.}(2010){Bond}, {O'Brien}, \&
  {Lauretta}}]{Bond-etal-2010}
{Bond}, J.~C., {O'Brien}, D.~P., \& {Lauretta}, D.~S. 2010, \apj, 715, 1050,
  \dodoi{10.1088/0004-637X/715/2/1050}

\bibitem[{{Borucki} {et~al.}(2010){Borucki}, {Koch}, {Basri}, {Batalha},
  {Brown}, {Caldwell}, {Caldwell}, {Christensen-Dalsgaard}, {Cochran},
  {DeVore}, {Dunham}, {Dupree}, {Gautier}, {Geary}, {Gilliland}, {Gould},
  {Howell}, {Jenkins}, {Kondo}, {Latham}, {Marcy}, {Meibom}, {Kjeldsen},
  {Lissauer}, {Monet}, {Morrison}, {Sasselov}, {Tarter}, {Boss}, {Brownlee},
  {Owen}, {Buzasi}, {Charbonneau}, {Doyle}, {Fortney}, {Ford}, {Holman},
  {Seager}, {Steffen}, {Welsh}, {Rowe}, {Anderson}, {Buchhave}, {Ciardi},
  {Walkowicz}, {Sherry}, {Horch}, {Isaacson}, {Everett}, {Fischer}, {Torres},
  {Johnson}, {Endl}, {MacQueen}, {Bryson}, {Dotson}, {Haas}, {Kolodziejczak},
  {Van Cleve}, {Chandrasekaran}, {Twicken}, {Quintana}, {Clarke}, {Allen},
  {Li}, {Wu}, {Tenenbaum}, {Verner}, {Bruhweiler}, {Barnes}, \&
  {Prsa}}]{Borucki2010}
{Borucki}, W.~J., {Koch}, D., {Basri}, G., {et~al.} 2010, Science, 327, 977,
  \dodoi{10.1126/science.1185402}

\bibitem[{Bouchet {et~al.}(2013)Bouchet, Mazevet, Morard, Guyot, \&
  Musella}]{bouchet}
Bouchet, J., Mazevet, S., Morard, G., Guyot, F., \& Musella, R. 2013, Physical
  Review B, 87, 094102

\bibitem[{{Bourrier} {et~al.}(2018){Bourrier}, {Dumusque}, {Dorn}, {Henry},
  {Astudillo-Defru}, {Rey}, {Benneke}, {H{\'e}brard}, {Lovis}, {Demory},
  {Moutou}, \& {Ehrenreich}}]{Bourrier+18}
{Bourrier}, V., {Dumusque}, X., {Dorn}, C., {et~al.} 2018, \aap, 619, A1,
  \dodoi{10.1051/0004-6361/201833154}

\bibitem[{{Boyajian} {et~al.}(2012{\natexlab{a}}){Boyajian}, {McAlister}, {van
  Belle}, {Gies}, {ten Brummelaar}, {von Braun}, {Farrington}, {Goldfinger},
  {O'Brien}, {Parks}, {Richardson}, {Ridgway}, {Schaefer}, {Sturmann},
  {Sturmann}, {Touhami}, {Turner}, \& {White}}]{Boyajian2012a}
{Boyajian}, T.~S., {McAlister}, H.~A., {van Belle}, G., {et~al.}
  2012{\natexlab{a}}, \apj, 746, 101, \dodoi{10.1088/0004-637X/746/1/101}

\bibitem[{{Boyajian} {et~al.}(2012{\natexlab{b}}){Boyajian}, {von Braun}, {van
  Belle}, {McAlister}, {ten Brummelaar}, {Kane}, {Muirhead}, {Jones}, {White},
  {Schaefer}, {Ciardi}, {Henry}, {L{\'o}pez-Morales}, {Ridgway}, {Gies}, {Jao},
  {Rojas-Ayala}, {Parks}, {Sturmann}, {Sturmann}, {Turner}, {Farrington},
  {Goldfinger}, \& {Berger}}]{Boyajian2012b}
{Boyajian}, T.~S., {von Braun}, K., {van Belle}, G., {et~al.}
  2012{\natexlab{b}}, \apj, 757, 112, \dodoi{10.1088/0004-637X/757/2/112}

\bibitem[{{Broeg} {et~al.}(2013){Broeg}, {Fortier}, {Ehrenreich}, {Alibert},
  {Baumjohann}, {Benz}, {Deleuil}, {Gillon}, {Ivanov}, {Liseau}, {Meyer},
  {Oloffson}, {Pagano}, {Piotto}, {Pollacco}, {Queloz}, {Ragazzoni}, {Renotte},
  {Steller}, \& {Thomas}}]{Broeg2013}
{Broeg}, C., {Fortier}, A., {Ehrenreich}, D., {et~al.} 2013, in European
  Physical Journal Web of Conferences, Vol.~47, European Physical Journal Web
  of Conferences, 3005

\bibitem[{{Cayrel} {et~al.}(1997){Cayrel}, {Castelli}, {Katz}, {van't Veer},
  {Gomez}, \& {Perrin}}]{Cayrel-etal-1997}
{Cayrel}, R., {Castelli}, F., {Katz}, D., {et~al.} 1997, in ESA Special
  Publication, Vol. 402, Hipparcos - Venice '97, ed. R.~M. {Bonnet},
  E.~{H{\o}g}, P.~L. {Bernacca}, L.~{Emiliani}, A.~{Blaauw}, C.~{Turon},
  J.~{Kovalevsky}, L.~{Lindegren}, H.~{Hassan}, M.~{Bouffard}, B.~{Strim},
  D.~{Heger}, M.~A.~C. {Perryman}, \& L.~{Woltjer}, 433--436

\bibitem[{Connolly(2009)}]{connolly09}
Connolly, J. 2009, Geochemistry, Geophysics, Geosystems, 10

\bibitem[{{Creevey} {et~al.}(2007){Creevey}, {Monteiro}, {Metcalfe}, {Brown},
  {Jim{\'e}nez-Reyes}, \& {Belmonte}}]{Creevey2007}
{Creevey}, O.~L., {Monteiro}, M.~J.~P.~F.~G., {Metcalfe}, T.~S., {et~al.} 2007,
  \apj, 659, 616, \dodoi{10.1086/512097}

\bibitem[{{Crida} {et~al.}(2018){Crida}, {Ligi}, {Dorn}, {Borsa}, \&
  {Lebreton}}]{Crida+2018b}
{Crida}, A., {Ligi}, R., {Dorn}, C., {Borsa}, F., \& {Lebreton}, Y. 2018,
  Research Notes of the American Astronomical Society, 2, 172,
  \dodoi{10.3847/2515-5172/aae1f6}

\bibitem[{{Daviau} \& {Lee}(2017{\natexlab{a}})}]{Daviau-Lee-2017b}
{Daviau}, K., \& {Lee}, K.~K.~M. 2017{\natexlab{a}}, \prb, 96, 174102,
  \dodoi{10.1103/PhysRevB.96.174102}

\bibitem[{{Daviau} \& {Lee}(2017{\natexlab{b}})}]{Daviau-Lee-2017a}
---. 2017{\natexlab{b}}, \prb, 95, 134108, \dodoi{10.1103/PhysRevB.95.134108}

\bibitem[{{Delgado Mena} {et~al.}(2010){Delgado Mena}, {Israelian},
  {Gonz{\'a}lez Hern{\'a}ndez}, {Bond}, {Santos}, {Udry}, \&
  {Mayor}}]{DelgadoMena-etal-2010}
{Delgado Mena}, E., {Israelian}, G., {Gonz{\'a}lez Hern{\'a}ndez}, J.~I.,
  {et~al.} 2010, \apj, 725, 2349, \dodoi{10.1088/0004-637X/725/2/2349}

\bibitem[{{Demory} {et~al.}(2012){Demory}, {Gillon}, {Seager}, {Benneke},
  {Deming}, \& {Jackson}}]{Demory-etal-2012}
{Demory}, B.-O., {Gillon}, M., {Seager}, S., {et~al.} 2012, \apjl, 751, L28,
  \dodoi{10.1088/2041-8205/751/2/L28}

\bibitem[{{Demory} {et~al.}(2011){Demory}, {Gillon}, {Deming}, {Valencia},
  {Seager}, {Benneke}, {Lovis}, {Cubillos}, {Harrington}, {Stevenson}, {Mayor},
  {Pepe}, {Queloz}, {S{\'e}gransan}, \& {Udry}}]{Demory2011}
{Demory}, B.-O., {Gillon}, M., {Deming}, D., {et~al.} 2011, \aap, 533, A114,
  \dodoi{10.1051/0004-6361/201117178}

\bibitem[{{Demory} {et~al.}(2016){Demory}, {Gillon}, {de Wit}, {Madhusudhan},
  {Bolmont}, {Heng}, {Kataria}, {Lewis}, {Hu}, {Krick}, {Stamenkovi{\'c}},
  {Benneke}, {Kane}, \& {Queloz}}]{Demory-etal-2016}
{Demory}, B.-O., {Gillon}, M., {de Wit}, J., {et~al.} 2016, \nat, 532, 207,
  \dodoi{10.1038/nature17169}

\bibitem[{Des~Etangs {et~al.}(2008)Des~Etangs, Pont, Vidal-Madjar, \&
  Sing}]{Lecavelier08}
Des~Etangs, A.~L., Pont, F., Vidal-Madjar, A., \& Sing, D. 2008, Astronomy \&
  Astrophysics, 481, L83

\bibitem[{{Dorn} \& {Heng}(2017)}]{Dorn-Heng-2017}
{Dorn}, C., \& {Heng}, K. 2017, ArXiv e-prints.
\newblock \doarXiv{1711.07745}

\bibitem[{{Dorn} {et~al.}(2017{\natexlab{a}}){Dorn}, {Hinkel}, \&
  {Venturini}}]{Dorn-etal-2017b}
{Dorn}, C., {Hinkel}, N.~R., \& {Venturini}, J. 2017{\natexlab{a}}, \aap, 597,
  A38, \dodoi{10.1051/0004-6361/201628749}

\bibitem[{{Dorn} {et~al.}(2015){Dorn}, {Khan}, {Heng}, {Connolly}, {Alibert},
  {Benz}, \& {Tackley}}]{Dorn-etal-2015}
{Dorn}, C., {Khan}, A., {Heng}, K., {et~al.} 2015, \aap, 577, A83,
  \dodoi{10.1051/0004-6361/201424915}

\bibitem[{{Dorn} {et~al.}(2017{\natexlab{b}}){Dorn}, {Venturini}, {Khan},
  {Heng}, {Alibert}, {Helled}, {Rivoldini}, \& {Benz}}]{Dorn-etal-2017a}
{Dorn}, C., {Venturini}, J., {Khan}, A., {et~al.} 2017{\natexlab{b}}, \aap,
  597, A37, \dodoi{10.1051/0004-6361/201628708}

\bibitem[{{Dragomir} {et~al.}(2014){Dragomir}, {Matthews}, {Winn}, \&
  {Rowe}}]{Dragomir-etal-2014}
{Dragomir}, D., {Matthews}, J.~M., {Winn}, J.~N., \& {Rowe}, J.~F. 2014, in IAU
  Symposium, Vol. 293, Formation, Detection, and Characterization of Extrasolar
  Habitable Planets, ed. N.~{Haghighipour}, 52--57

\bibitem[{{Ehrenreich} {et~al.}(2012){Ehrenreich}, {Bourrier}, {Bonfils},
  {Lecavelier des Etangs}, {H{\'e}brard}, {Sing}, {Wheatley}, {Vidal-Madjar},
  {Delfosse}, {Udry}, {Forveille}, \& {Moutou}}]{Ehrenreich-etal-2012}
{Ehrenreich}, D., {Bourrier}, V., {Bonfils}, X., {et~al.} 2012, \aap, 547, A18,
  \dodoi{10.1051/0004-6361/201219981}

\bibitem[{{Endl} {et~al.}(2012){Endl}, {Robertson}, {Cochran}, {MacQueen},
  {Brugamyer}, {Caldwell}, {Wittenmyer}, {Barnes}, \&
  {Gullikson}}]{Endl-etal-2012}
{Endl}, M., {Robertson}, P., {Cochran}, W.~D., {et~al.} 2012, \apj, 759, 19,
  \dodoi{10.1088/0004-637X/759/1/19}

\bibitem[{{Fischer} {et~al.}(2008){Fischer}, {Marcy}, {Butler}, {Vogt},
  {Laughlin}, {Henry}, {Abouav}, {Peek}, {Wright}, {Johnson}, {McCarthy}, \&
  {Isaacson}}]{Fischer-etal-2008}
{Fischer}, D.~A., {Marcy}, G.~W., {Butler}, R.~P., {et~al.} 2008, \apj, 675,
  790, \dodoi{10.1086/525512}

\bibitem[{{Flower}(1996)}]{Flower-1996}
{Flower}, P.~J. 1996, \apj, 469, 355, \dodoi{10.1086/177785}

\bibitem[{Gordon \& McBride(1994)}]{CEA}
Gordon, S., \& McBride, B.~J. 1994, Computer program for calculation of complex
  chemical equilibrium compositions and applications, Vol.~1 (National
  Aeronautics and Space Administration, Office of Management, Scientific and
  Technical Information Program)

\bibitem[{{Guillot}(2010)}]{Guillot-2010}
{Guillot}, T. 2010, \aap, 520, A27, \dodoi{10.1051/0004-6361/200913396}

\bibitem[{Heng {et~al.}(2014)Heng, Mendon{\c{c}}a, \& Lee}]{Heng2014}
Heng, K., Mendon{\c{c}}a, J.~M., \& Lee, J.-M. 2014, The Astrophysical Journal
  Supplement Series, 215, 4

\bibitem[{{Hinkel} {et~al.}(2014){Hinkel}, {Timmes}, {Young}, {Pagano}, \&
  {Turnbull}}]{Hinkel-etal-2014}
{Hinkel}, N.~R., {Timmes}, F.~X., {Young}, P.~A., {Pagano}, M.~D., \&
  {Turnbull}, M.~C. 2014, \aj, 148, 54, \dodoi{10.1088/0004-6256/148/3/54}

\bibitem[{Jin {et~al.}(2014)Jin, Mordasini, Parmentier, Van~Boekel, Henning, \&
  Ji}]{JIN2014}
Jin, S., Mordasini, C., Parmentier, V., {et~al.} 2014, The Astrophysical
  Journal, 795, 65

\bibitem[{{Kervella} {et~al.}(2004){Kervella}, {Th{\'e}venin}, {Di Folco}, \&
  {S{\'e}gransan}}]{Kervella2004}
{Kervella}, P., {Th{\'e}venin}, F., {Di Folco}, E., \& {S{\'e}gransan}, D.
  2004, \aap, 426, 297, \dodoi{10.1051/0004-6361:20035930}

\bibitem[{{Kuchner} \& {Seager}(2005)}]{Kuchner-Seager2005}
{Kuchner}, M.~J., \& {Seager}, S. 2005, ArXiv Astrophysics e-prints

\bibitem[{{Lebreton}(2012)}]{Lebreton2012}
{Lebreton}, Y. 2012, in Astronomical Society of the Pacific Conference Series,
  Vol. 462, Progress in Solar/Stellar Physics with Helio- and Asteroseismology,
  ed. H.~{Shibahashi}, M.~{Takata}, \& A.~E. {Lynas-Gray}, 469

\bibitem[{{Lebreton} \& {Goupil}(2014)}]{Lebreton-Goupil-2014}
{Lebreton}, Y., \& {Goupil}, M.~J. 2014, \aap, 569, A21,
  \dodoi{10.1051/0004-6361/201423797}

\bibitem[{{Ligi}(2014)}]{Ligi2014}
{Ligi}, R. 2014, in EAS Publications Series, Vol.~69, EAS Publications Series,
  273--283

\bibitem[{{Ligi} {et~al.}(2015){Ligi}, {Mourard}, {Lagrange}, {Perraut}, \&
  {Chiavassa}}]{Ligi-etal-2015}
{Ligi}, R., {Mourard}, D., {Lagrange}, A.-M., {Perraut}, K., \& {Chiavassa}, A.
  2015, \aap, 574, A69, \dodoi{10.1051/0004-6361/201424013}

\bibitem[{{Ligi} {et~al.}(2012){Ligi}, {Mourard}, {Lagrange}, {Perraut},
  {Boyajian}, {B{\'e}rio}, {Nardetto}, {Tallon-Bosc}, {McAlister}, {ten
  Brummelaar}, {Ridgway}, {Sturmann}, {Sturmann}, {Turner}, {Farrington}, \&
  {Goldfinger}}]{Ligi-etal-2012}
{Ligi}, R., {Mourard}, D., {Lagrange}, A.~M., {et~al.} 2012, \aap, 545, A5,
  \dodoi{10.1051/0004-6361/201219467}

\bibitem[{{Ligi} {et~al.}(2016){Ligi}, {Creevey}, {Mourard}, {Crida},
  {Lagrange}, {Nardetto}, {Perraut}, {Schultheis}, {Tallon-Bosc}, \& {ten
  Brummelaar}}]{Ligi-etal-2016}
{Ligi}, R., {Creevey}, O., {Mourard}, D., {et~al.} 2016, \aap, 586, A94,
  \dodoi{10.1051/0004-6361/201527054}

\bibitem[{{Lopez}(2017)}]{Lopez-2017}
{Lopez}, E.~D. 2017, \mnras, 472, 245, \dodoi{10.1093/mnras/stx1558}

\bibitem[{{Madhusudhan} {et~al.}(2012){Madhusudhan}, {Lee}, \&
  {Mousis}}]{Madhusudhan-etal-2012}
{Madhusudhan}, N., {Lee}, K.~K.~M., \& {Mousis}, O. 2012, \apjl, 759, L40,
  \dodoi{10.1088/2041-8205/759/2/L40}

\bibitem[{{Maxted} {et~al.}(2015){Maxted}, {Serenelli}, \&
  {Southworth}}]{Maxted-etal-2015_mass-age}
{Maxted}, P.~F.~L., {Serenelli}, A.~M., \& {Southworth}, J. 2015, \aap, 575,
  A36, \dodoi{10.1051/0004-6361/201425331}

\bibitem[{{Mayor} \& {Queloz}(1995)}]{Mayor1995}
{Mayor}, M., \& {Queloz}, D. 1995, \nat, 378, 355, \dodoi{10.1038/378355a0}

\bibitem[{{Mayor} {et~al.}(2003){Mayor}, {Pepe}, {Queloz}, {Bouchy},
  {Rupprecht}, {Lo Curto}, {Avila}, {Benz}, {Bertaux}, {Bonfils}, {Dall},
  {Dekker}, {Delabre}, {Eckert}, {Fleury}, {Gilliotte}, {Gojak}, {Guzman},
  {Kohler}, {Lizon}, {Longinotti}, {Lovis}, {Megevand}, {Pasquini}, {Reyes},
  {Sivan}, {Sosnowska}, {Soto}, {Udry}, {van Kesteren}, {Weber}, \&
  {Weilenmann}}]{Mayor-etal-2003}
{Mayor}, M., {Pepe}, F., {Queloz}, D., {et~al.} 2003, The Messenger, 114, 20

\bibitem[{{Morel} \& {Lebreton}(2008)}]{Morel-Lebreton-2008}
{Morel}, P., \& {Lebreton}, Y. 2008, \apss, 316, 61,
  \dodoi{10.1007/s10509-007-9663-9}

\bibitem[{{Moriarty} {et~al.}(2014){Moriarty}, {Madhusudhan}, \&
  {Fischer}}]{Moriatry-etal-2014}
{Moriarty}, J., {Madhusudhan}, N., \& {Fischer}, D. 2014, \apj, 787, 81,
  \dodoi{10.1088/0004-637X/787/1/81}

\bibitem[{{Mourard} {et~al.}(2009){Mourard}, {Clausse}, {Marcotto}, {Perraut},
  {Tallon-Bosc}, {B{\'e}rio}, {Blazit}, {Bonneau}, {Bosio}, {Bresson},
  {Chesneau}, {Delaa}, {H{\'e}nault}, {Hughes}, {Lagarde}, {Merlin}, {Roussel},
  {Spang}, {Stee}, {Tallon}, {Antonelli}, {Foy}, {Kervella}, {Petrov},
  {Thiebaut}, {Vakili}, {McAlister}, {ten Brummelaar}, {Sturmann}, {Sturmann},
  {Turner}, {Farrington}, \& {Goldfinger}}]{Mourard-etal-2009}
{Mourard}, D., {Clausse}, J.~M., {Marcotto}, A., {et~al.} 2009, \aap, 508,
  1073, \dodoi{10.1051/0004-6361/200913016}

\bibitem[{{Nisr} {et~al.}(2017){Nisr}, {Meng}, {MacDowell}, {Yan},
  {Prakapenka}, \& {Shim}}]{Nisr-et-al-2017}
{Nisr}, C., {Meng}, Y., {MacDowell}, A.~A., {et~al.} 2017, Journal of
  Geophysical Research (Planets), 122, 124, \dodoi{10.1002/2016JE005158}

\bibitem[{{Ram{\'{\i}}rez} {et~al.}(2014){Ram{\'{\i}}rez}, {Mel{\'e}ndez},
  {Bean}, {Asplund}, {Bedell}, {Monroe}, {Casagrande}, {Schirbel}, {Dreizler},
  {Teske}, {Tucci Maia}, {Alves-Brito}, \& {Baumann}}]{Ramirez-etal-2014}
{Ram{\'{\i}}rez}, I., {Mel{\'e}ndez}, J., {Bean}, J., {et~al.} 2014, \aap, 572,
  A48, \dodoi{10.1051/0004-6361/201424244}

\bibitem[{{Rauer} {et~al.}(2014){Rauer}, {Catala}, {Aerts}, {Appourchaux},
  {Benz}, {Brandeker}, {Christensen-Dalsgaard}, {Deleuil}, {Gizon}, {Goupil},
  {G{\"u}del}, {Janot-Pacheco}, {Mas-Hesse}, {Pagano}, {Piotto}, {Pollacco},
  {Santos}, {Smith}, {Su{\'a}rez}, {Szab{\'o}}, {Udry}, {Adibekyan}, {Alibert},
  {Almenara}, {Amaro-Seoane}, {Eiff}, {Asplund}, {Antonello}, {Barnes},
  {Baudin}, {Belkacem}, {Bergemann}, {Bihain}, {Birch}, {Bonfils}, {Boisse},
  {Bonomo}, {Borsa}, {Brand{\~a}o}, {Brocato}, {Brun}, {Burleigh}, {Burston},
  {Cabrera}, {Cassisi}, {Chaplin}, {Charpinet}, {Chiappini}, {Church},
  {Csizmadia}, {Cunha}, {Damasso}, {Davies}, {Deeg}, {D{\'{\i}}az}, {Dreizler},
  {Dreyer}, {Eggenberger}, {Ehrenreich}, {Eigm{\"u}ller}, {Erikson}, {Farmer},
  {Feltzing}, {de Oliveira Fialho}, {Figueira}, {Forveille}, {Fridlund},
  {Garc{\'{\i}}a}, {Giommi}, {Giuffrida}, {Godolt}, {Gomes da Silva},
  {Granzer}, {Grenfell}, {Grotsch-Noels}, {G{\"u}nther}, {Haswell}, {Hatzes},
  {H{\'e}brard}, {Hekker}, {Helled}, {Heng}, {Jenkins}, {Johansen},
  {Khodachenko}, {Kislyakova}, {Kley}, {Kolb}, {Krivova}, {Kupka}, {Lammer},
  {Lanza}, {Lebreton}, {Magrin}, {Marcos-Arenal}, {Marrese}, {Marques},
  {Martins}, {Mathis}, {Mathur}, {Messina}, {Miglio}, {Montalban}, {Montalto},
  {Monteiro}, {Moradi}, {Moravveji}, {Mordasini}, {Morel}, {Mortier},
  {Nascimbeni}, {Nelson}, {Nielsen}, {Noack}, {Norton}, {Ofir}, {Oshagh},
  {Ouazzani}, {P{\'a}pics}, {Parro}, {Petit}, {Plez}, {Poretti}, {Quirrenbach},
  {Ragazzoni}, {Raimondo}, {Rainer}, {Reese}, {Redmer}, {Reffert},
  {Rojas-Ayala}, {Roxburgh}, {Salmon}, {Santerne}, {Schneider}, {Schou},
  {Schuh}, {Schunker}, {Silva-Valio}, {Silvotti}, {Skillen}, {Snellen}, {Sohl},
  {Sousa}, {Sozzetti}, {Stello}, {Strassmeier}, {{\v S}vanda}, {Szab{\'o}},
  {Tkachenko}, {Valencia}, {Van Grootel}, {Vauclair}, {Ventura}, {Wagner},
  {Walton}, {Weingrill}, {Werner}, {Wheatley}, \& {Zwintz}}]{Rauer-etal-2014}
{Rauer}, H., {Catala}, C., {Aerts}, C., {et~al.} 2014, Experimental Astronomy,
  38, 249, \dodoi{10.1007/s10686-014-9383-4}

\bibitem[{{Ricker} {et~al.}(2014){Ricker}, {Winn}, {Vanderspek}, {Latham},
  {Bakos}, {Bean}, {Berta-Thompson}, {Brown}, {Buchhave}, {Butler}, {Butler},
  {Chaplin}, {Charbonneau}, {Christensen-Dalsgaard}, {Clampin}, {Deming},
  {Doty}, {De Lee}, {Dressing}, {Dunham}, {Endl}, {Fressin}, {Ge}, {Henning},
  {Holman}, {Howard}, {Ida}, {Jenkins}, {Jernigan}, {Johnson}, {Kaltenegger},
  {Kawai}, {Kjeldsen}, {Laughlin}, {Levine}, {Lin}, {Lissauer}, {MacQueen},
  {Marcy}, {McCullough}, {Morton}, {Narita}, {Paegert}, {Palle}, {Pepe},
  {Pepper}, {Quirrenbach}, {Rinehart}, {Sasselov}, {Sato}, {Seager},
  {Sozzetti}, {Stassun}, {Sullivan}, {Szentgyorgyi}, {Torres}, {Udry}, \&
  {Villasenor}}]{Ricker2014}
{Ricker}, G.~R., {Winn}, J.~N., {Vanderspek}, R., {et~al.} 2014, in Society of
  Photo-Optical Instrumentation Engineers (SPIE) Conference Series, Vol. 9143,
  20

\bibitem[{{Seager} \& {Mall{\'e}n-Ornelas}(2003)}]{Seager-MallenOrnelas-2003}
{Seager}, S., \& {Mall{\'e}n-Ornelas}, G. 2003, \apj, 585, 1038,
  \dodoi{10.1086/346105}

\bibitem[{{Teske} {et~al.}(2013){Teske}, {Cunha}, {Schuler}, {Griffith}, \&
  {Smith}}]{Teske-etal-2013}
{Teske}, J.~K., {Cunha}, K., {Schuler}, S.~C., {Griffith}, C.~A., \& {Smith},
  V.~V. 2013, \apj, 778, 132, \dodoi{10.1088/0004-637X/778/2/132}

\bibitem[{{Torres}(2010)}]{Torres-2010}
{Torres}, G. 2010, \aj, 140, 1158, \dodoi{10.1088/0004-6256/140/5/1158}

\bibitem[{{Tsiaras} {et~al.}(2016){Tsiaras}, {Rocchetto}, {Waldmann}, {Venot},
  {Varley}, {Morello}, {Damiano}, {Tinetti}, {Barton}, {Yurchenko}, \&
  {Tennyson}}]{Tsiaras-etal-2016}
{Tsiaras}, A., {Rocchetto}, M., {Waldmann}, I.~P., {et~al.} 2016, \apj, 820,
  99, \dodoi{10.3847/0004-637X/820/2/99}

\bibitem[{{von Braun} {et~al.}(2011){von Braun}, {Boyajian}, {ten Brummelaar},
  {Kane}, {van Belle}, {Ciardi}, {Raymond}, {L{\'o}pez-Morales}, {McAlister},
  {Schaefer}, {Ridgway}, {Sturmann}, {Sturmann}, {White}, {Turner},
  {Farrington}, \& {Goldfinger}}]{vanBraun2011}
{von Braun}, K., {Boyajian}, T.~S., {ten Brummelaar}, T.~A., {et~al.} 2011,
  \apj, 740, 49, \dodoi{10.1088/0004-637X/740/1/49}

\bibitem[{{Wilson} \& {Militzer}(2014)}]{Wilson-and-Militzer-2014}
{Wilson}, H.~F., \& {Militzer}, B. 2014, \apj, 793, 34,
  \dodoi{10.1088/0004-637X/793/1/34}

\bibitem[{{Winn} {et~al.}(2011){Winn}, {Matthews}, {Dawson}, {Fabrycky},
  {Holman}, {Kallinger}, {Kuschnig}, {Sasselov}, {Dragomir}, {Guenther},
  {Moffat}, {Rowe}, {Rucinski}, \& {Weiss}}]{Winn-etal-2011}
{Winn}, J.~N., {Matthews}, J.~M., {Dawson}, R.~I., {et~al.} 2011, \apjl, 737,
  L18, \dodoi{10.1088/2041-8205/737/1/L18}

\end{thebibliography}



\end{document}